\documentclass[10pt,a4paper]{article}%
\usepackage{natbib}
\usepackage{a4wide}

\usepackage{mathpazo}
\usepackage[T1]{fontenc}

\usepackage{amsmath}
\usepackage{amsfonts}
\usepackage{amssymb}
\usepackage{amsthm}
\usepackage{enumerate}

\usepackage{microtype}

\usepackage{graphicx}
	\setkeys{Gin}{width=1\textwidth}

\numberwithin{equation}{section}

\usepackage[english]{babel}
\usepackage[latin1]{inputenc}
\usepackage{mathrsfs}

\usepackage[x11names]{xcolor}

\usepackage{hyperref}
	\hypersetup{pdfauthor=Tristan LAUNAY,
				pdfstartview=FitH, pdfstartpage=1, bookmarksopen=true,
		    bookmarksopenlevel=0, bookmarksnumbered=true, naturalnames=true, breaklinks=false, linktocpage=true}
\usepackage[all]{hypcap}

\usepackage{dsfont}

\usepackage{etex}
\usepackage{pgfplots,pgfplotstable,booktabs,colortbl,array}
\pgfkeys{
/pgf/number format/1000 sep={\;},
/pgf/number format/.cd, 
/pgf/number format/precision=2,
/pgfplots/table/every odd row/.style={before row={\rowcolor[gray]{1.0}}},
/pgfplots/table/every even row/.style={before row={\rowcolor[gray]{0.9}}}, 
/pgfplots/table/every head row/.style={before row={\rowcolor[gray]{1.0}\toprule},after row=\bottomrule},
/pgfplots/table/every last row/.style={after row=\toprule}
}
\usepackage{tikz}
\usepackage{calc}

\usepackage{xspace}

\newcommand{\N}{\mathbb{N}}

\newcommand{\R}{\mathbb{R}}

\newcommand{\one}{\mathds{1}}

\newcommand{\esp}{\mathbb{E}}

\DeclareMathOperator{\var}{Var}
\newcommand{\proba}{\mathbb{P}}
\newcommand{\loi}[1]{\mathcal{#1}}

\newcommand{\nud}{\mathrm{d}}
\newcommand{\ud}{\,\mathrm{d}}

\newcommand{\interoo}[2]{]#1,\,#2[}
\newcommand{\interof}[2]{]#1,\,#2]}

\usepackage{framed}
\definecolor{shadecolor}{gray}{0.9}
\newenvironment{algorithm}[1]{\begin{shaded}\begin{algo}[#1]\upshape}{\end{algo}\end{shaded}}

\newtheorem{algo}{\iflanguage{french}{Algorithme}{Algorithm}}[section]

\begin{document}
\setcounter{secnumdepth}{2}

\title{On particle filters applied to electricity load forecasting}
\author{Tristan Launay\(^{1,2}\) \and Anne Philippe\(^1\) \and Sophie Lamarche\(^2\)}
\maketitle
\footnotetext[1]{Laboratoire de Math\'ematiques Jean Leray, 2 Rue de la Houssini\`ere -- BP 92208, 44322 Nantes Cedex 3, France}
\footnotetext[2]{Electricit\'e de France R\&D, 1 Avenue du G\'en\'eral de Gaulle, 92141 Clamart Cedex, France}
\begin{abstract}
In this paper, we are interested in the online prediction of the electricity load, within the Bayesian framework of dynamic models. We offer a review of sequential Monte Carlo methods, and provide the calculations needed for the derivation of so-called particles filters. We also discuss the practical issues arising from their use, and some of the variants proposed in the literature to deal with them, giving detailed algorithms whenever possible for an easy implementation. We propose an additional step to help make basic particle filters more robust with regard to outlying observations. Finally we use such a particle filter to estimate a state-space model that includes exogenous variables in order to forecast the electricity load for the customers of the French electricity company \'Electricit\'e de France and discuss the various results obtained.
\medskip

\end{abstract}


\section{Introduction}\label{sec:intro3}
Let \(\{X_n\}_{n\geq 0}\) and \(\{Y_n\}_{n\geq 0}\) be \(\mathcal{X}\subset\R^{n_x}\) and \(\mathcal{Y}\subset\R^{n_y}\)-valued stochastic processes defined on a measurable space. The observations \(\{Y_n\}_{n\geq 0}\) are assumed conditionally independent given the hidden Markov process \(\{X_n\}_{n\geq 0}\) most often referred to as the states of the model, and are characterised by the conditional density \(g_n^\theta(y_n|x_n)\). We denote the initial density of the state as \(\mu^\theta(x_0)\) and the Markov transition density from time \(n-1\) to time \(n\) as \(f_n^\theta(x_n|x_{n-1})\). The superscript \(\theta\) on these densities is the parameter of the model, that belongs to an open set \(\Theta\subset\R^{n_\theta}\). The model can be summarised (using practical and common if not exactly rigorous notations) as
\begin{align}
X_0\sim \mu^\theta(\cdot),\quad X_n|(X_{n-1}=x_{n-1})&\sim f_{n}^\theta(\cdot|x_{n-1}) \label{eq:dynamicprior}\\
Y_n|(X_n=x_n) &\sim g_n^\theta(\cdot|x_n). \label{eq:dynamicobservations}
\end{align}
Within the Bayesian framework, equations \eqref{eq:dynamicprior} specify the prior on the states of the model whose likelihood is defined via \eqref{eq:dynamicobservations}.

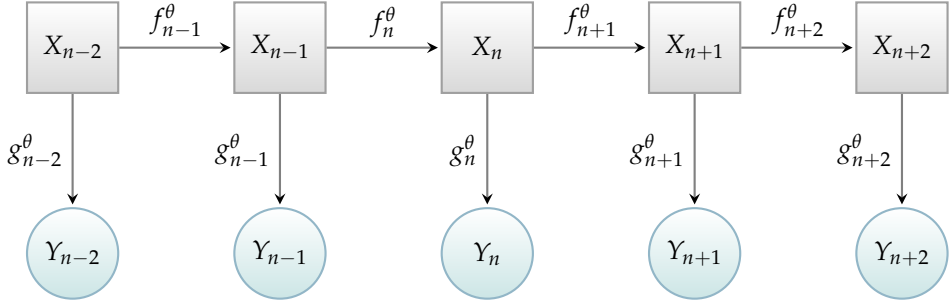
\begin{figure}[htbp]
\begin{center}
\begin{tikzpicture}[
  point/.style={coordinate},
  >=stealth,thick,draw=black!50,
  tip/.style={->,shorten >=1pt},
  hv path/.style={to path={-| (\tikztotarget)}},
  vh path/.style={to path={|- (\tikztotarget)}},
  state/.style={rectangle, minimum height=12mm, minimum width=12mm, draw=gray!50!black!50,   top color=white, bottom color=gray!50!black!20,    font=\normalfont},
  obs/.style={circle,   minimum height=12mm, minimum width=12mm, draw=cyan!50!black!50,   top color=white, bottom color=cyan!50!black!20,  font=\normalfont},
  ]
  \matrix[row sep=1.5cm,column sep=1.5cm,ampersand replacement=\&]{
       \node (xm2) [state]  {\(X_{n-2}\)};
    \& \node (xm1) [state]  {\(X_{n-1}\)};
    \& \node (x)   [state]  {\(X_{n}\)  };
    \& \node (xp1) [state]  {\(X_{n+1}\)};
    \& \node (xp2) [state]  {\(X_{n+2}\)}; \\
       \node (ym2) [obs]    {\(Y_{n-2}\)};
    \& \node (ym1) [obs]    {\(Y_{n-1}\)};
    \& \node (y)   [obs]    {\(Y_{n}\)  };
    \& \node (yp1) [obs]    {\(Y_{n+1}\)};
    \& \node (yp2) [obs]    {\(Y_{n+2}\)}; \\
    };
  \path
        (xm2)   edge[tip] node[above] {\(f_{n-1}^\theta\)} (xm1)
        (xm1)   edge[tip] node[above] {\(f_{n  }^\theta\)} (x)
        (x)     edge[tip] node[above] {\(f_{n+1}^\theta\)} (xp1)
        (xp1)   edge[tip] node[above] {\(f_{n+2}^\theta\)} (xp2)

        (xm2)   edge[tip] node[left] {\(g_{n-2}^\theta\)} (ym2)
        (xm1)   edge[tip] node[left] {\(g_{n-1}^\theta\)} (ym1)
        (x)     edge[tip] node[left] {\(g_{n  }^\theta\)} (y)
        (xp1)   edge[tip] node[left] {\(g_{n+1}^\theta\)} (yp1)
        (xp2)   edge[tip] node[left] {\(g_{n+2}^\theta\)} (yp2) ;
\end{tikzpicture}
\end{center}
 \caption{A generic hidden Markov Model (HMM).}
 \label{fig:schemaHMM}
\end{figure}

Notice here that we restrict ourselves to models with independent observations, but that the framework can easily be extended to include dependent observations if need be. The class of dynamic models we consider, known as general state-space models or hidden Markov models (HMM) in the literature and whose typical representation is given in Figure \ref{fig:schemaHMM}, includes many non linear and non Gaussian time series models such as
\begin{align}
X_{n+1} &= F_n(X_n, V_{n+1})\\
Y_n &= G_n(X_n,W_n)
\end{align}
where \(\{V_n\}_{n\geq1}\) and \(\{W_n\}_{n\geq0}\) are independent sequences of independent random variables and \(\{F_n\}_{n\geq1}\) and \(\{G_n\}_{n\geq1}\) are sequences of (possibly non linear) functions. Such models find applications in many fields including time-series forecasting \citep{Dordonnat2}, biostatistics \citep{Rossi,Vavoulis2012}, econometrics \citep{LiuWest,Johansen,ChopinJacob2012}, telecommunications \citep{Lee2010}, object tracking \citep{Rui2001,Gilks,Gustafsson,Karlsson2005}, etc.

When the parameter \(\theta\) is known, on-line inference about the state process given the observations is a so-called optimal filtering problem. For simple models such as the linear Gaussian state-space model the problem can be solved exactly using the standard Kalman filter \cite[see for example][]{Durbin}, and the case of a finite state-space also allows for explicit calculations. For non linear models, the Extended Kalman filter is often used and relies on the approximation of the first derivative of \(F_n\), although good performances are not guaranteed theoretically. Another technique is the so-called Unscented Kalman filter \cite[see][for the comprehensive details]{Wan2000Unscented} which makes use of the unscented transformation to deal with the non linearity of the system.

For our application, we are interested in the on-line prediction of the french electricity load through the estimation (and prediction) of a dynamic model and choose to consider Sequential Monte Carlo (SMC) methods also known as particle methods instead. SMC methods are a class of sequential simulation-based algorithms which aim at approximating the posterior distributions of interest. They represent a popular alternative to Kalman filters \citep{Kantas} since they are often easy to implement, apply to non linear non Gaussian models, and have been demonstrated to yield accurate estimates \citep{DoucetDeFreitas,Liu2004}.

In Section \ref{sec:towardsSMC} we introduce the key concepts behind sequential Monte Carlo methods. In Section \ref{sec:particlefilters} we first derive the algorithm for a basic particle filter and discuss common practical issues. We then review the main techniques appearing in the literature to deal with these issues and we also propose a new additional step to help make particle filters more robust with regard to outlying observations. Finally, we propose a new nonlinear dynamic model for the electricity load in Section \ref{sec:applicationdynamic} and use a particle filter to estimate this model. We compare the predictions we obtain to operational predictions and show that our model remains competitive, even though its definition is simpler than that of the model studied in \cite{Dordonnat}.

\section{Inference in hidden Markov models}\label{sec:towardsSMC}
Let us first assume that the parameter \(\theta\) is known: the model with \(\theta\) unknown will be discussed later in Section \ref{subsec:parameterestimation}. Given equations \eqref{eq:dynamicprior} and \eqref{eq:dynamicobservations}, the posterior distribution of the states given the observations is
\begin{align}
  \pi^\theta(x_{0:n}|y_{0:n}) \propto \underbrace{\prod_{k=1}^n g_k^\theta(y_k|x_k)}_{n \text{ likelihoods}} \cdot \underbrace{\prod_{k=1}^n f_k^\theta(x_k|x_{k-1})}_{n \text{ transition densities}} \cdot \underbrace{\vphantom{\prod_{k=1}^n g_k^\theta(y_k|x_k)}\mu^\theta(x_0)}_{\text{initial density}}. \label{eq:dynamicposterior}
  \end{align}
From equation \eqref{eq:dynamicposterior}, three distinct goals might be pursued \cite[see for example][]{Chen,CappeMoulines}
\begin{description}
 \item[Filtering:] the aim of filtering is to estimate the distribution of the state \(X_n\) conditionally to the observations up to time \(n\), i.e. \(y_{0:n}\).
 \item[Smoothing:] the aim of smoothing is to estimate the distribution of the state \(X_n\) conditionally to the observations up to time \(n^\prime\) (with \(n^\prime \geq n\)), i.e. \(y_{0:{n^\prime}}\). Note that \(\pi^\theta(x_n|y_{0:n})\) is both a filtered and a smoothed distribution.
 \item[Predicting:] the aim of predicting is to estimate the distribution of the state \(X_{n+\tau}\) (with an horizon \(\tau > 0\)) conditionally to the observations up to time \(n\), i.e. \(y_{0:n}\). From there, using \eqref{eq:dynamicobservations}, it is easy to forecast the upcoming observation \(Y_{n+\tau}\) which is usually the real target. When not explicitly mentioned, the horizon considered for prediction will be \(\tau = 1\).
\end{description}
To summarise, given the available observations, filtering focuses on the current state, smoothing focuses on the past states, and predicting focuses on the future states. Our goal being the online prediction of the electricity load, we chose to focus on predicting and filtering, since the filtered distribution of the state at time \(n\) is needed to produce forecasts for time \(n+\tau\): ultimately, smoothing only refines the estimation of past states over time, without influencing the quality of the online prediction, and is therefore not needed to achieve our goal.

\subsection*{Markov Chains Monte Carlo}
MCMC methods \cite[see for example][]{Robert2,RobertCasella,MarinRobert} certainly represent a viable estimation procedure: most of the time, nothing really prevents the exploration via MCMC of the posterior distribution derived in \eqref{eq:dynamicposterior} from the prior and the likelihood given in \eqref{eq:dynamicprior} and \eqref{eq:dynamicobservations}. From a practical point of view however, MCMC methods are most likely not the optimal tool: the addition of a new observation \(y_{n+1}\) from the model forces the overall re-estimation of the smoothed distribution of the states \(\pi^\theta(x_{0:n+1}|y_{0:n+1})\) even when we are interested only in the last marginal of this distribution i.e. the filtered distribution \(\pi^\theta(x_{n+1}|y_{0:n+1})\). The MCMC estimation is thus not recursive (with regard to the time index) in the sense that the filtered distribution \(\pi^\theta(x_{n+1}|y_{0:n+1})\) at time \(n+1\) cannot be computed from the previous filtered distribution \(\pi^\theta(x_{n}|y_{0:n})\) at time \(n\) using MCMC methods, which is a major drawback given the computationally expensive nature of these methods.

Notice also that even though designing the MCMC algorithm can be simple in some cases, the dimension of the space explored grows linearly with the time index making the assessment of the convergence of the produced Markov chains all the more complicated.

\subsection*{Importance sampling}
Monte Carlo integration allows the estimation of integrals of the form
\begin{align}
I &= \esp^\pi[h(X)] = \int h(x) \pi(x) \ud x, \label{eq:aimMCintegration}
\end{align}
where \(\pi\) is a probability density and where \(h\in \mathrm{L^1}(\pi)\). This method is often used to numerically approximate the expectation of a random variable whose density is \(\pi\) or a moment of higher order.

Let us assume that a probability density \(q\) (the so-called importance density) is available from which we can simulate, and such that the support of \(\pi\) is included in that of \(q\). We can then write
\begin{align*}
I &= \int h(x) \pi(x) \ud x = \dfrac{\displaystyle \int \dfrac{h(x) \pi(x)}{q(x)} q(x) \ud x}{\displaystyle \int \dfrac{\pi(x)}{q(x)} q(x) \ud x}.
\end{align*}
Given \(X^1, \ldots, X^M\) i.i.d. random variables with probability density \(q\), the self-normalised importance sampling estimator of \(I\) is defined by
\begin{align}
\widehat{I}_M(q)
&= \frac{\displaystyle\sum_{j=1}^M \dfrac{h(X^j)\pi(X^j)}{q(X^j)}}{\displaystyle\sum_{j=1}^M \dfrac{\pi(X^j)}{q(X^j)}}
= \sum_{j=1}^M w^j h(X^j),\label{eq:estimateur.IS2}
\end{align}
where we define the self-normalised weights as
\begin{align}
w^j &= \frac{\widetilde{w}^j}{\sum_{k=1}^M \widetilde{w}^k}.\label{eq:definitionpoids}
\end{align}
with
\begin{align}
\widetilde{w}^j &= \dfrac{\pi(X^j)}{q(X^j)}.\label{eq:definitionpoidstilde}
\end{align}
See \cite{Geweke1989Bayesian} for the theoretical details (including proof of the consistency of the estimator).

\subsection*{Sequential Monte Carlo}
SMC methods provide a viable and popular alternative to MCMC methods for the Bayesian online estimation of dynamic models. Particle methods are recursive by nature (thus computationally cheaper than MCMC) and similar in some ways to the Kalman filter approach. Particle methods essentially draw their strength from the immediate calculations that we show below
\begin{align*}
\pi^\theta(x_{0:n}|y_{0:n})
&= \dfrac{\pi^\theta(y_{0:n}|x_{0:n})\pi^\theta(x_{0:n})}{\pi^\theta(y_{0:n})}
= \dfrac{\pi^\theta(y_n, y_{0:{n-1}}|x_{0:n})\pi^\theta(x_{0:n})}{\pi^\theta(y_n, y_{0:{n-1}})} \\
&= \dfrac{\pi^\theta(y_n | y_{0:{n-1}}, x_{0:n})\pi^\theta(y_{0:{n-1}} | x_{0:n})\pi^\theta(x_{0:n})}{\pi^\theta(y_n | y_{0:{n-1}})\pi^\theta(y_{0:{n-1}})} \\
&= \dfrac{\pi^\theta(y_n | y_{0:{n-1}}, x_{0:n})\pi^\theta(x_{0:n} | y_{0:{n-1}})\pi^\theta(y_{0:{n-1}})\pi^\theta(x_{0:n})}{\pi^\theta(y_n | y_{0:{n-1}})\pi^\theta(y_{0:{n-1}})\pi^\theta(x_{0:n})} \\
&= \dfrac{\pi^\theta(y_n | x_{0:n})\pi^\theta(x_n|x_{0:{n-1}}, y_{0:{n-1}})}{\pi^\theta(y_n | y_{0:{n-1}})} \cdot \pi^\theta(x_{0:n-1}|y_{0:{n-1}})
\end{align*}
i.e. with the notations we introduced earlier:
\begin{align}
\pi^\theta(x_{0:n}|y_{0:n}) &= \dfrac{g_n^\theta(y_n | x_n)f_n^\theta(x_n|x_{n-1})}{\pi^\theta(y_n | y_{0:{n-1}})} \cdot \pi^\theta(x_{0:n-1}|y_{0:{n-1}}) \label{eq:recursion} \\
&\propto g_n^\theta(y_n | x_n)f_n^\theta(x_n|x_{n-1}) \cdot \pi^\theta(x_{0:n-1}|y_{0:{n-1}}). \nonumber
\end{align}
The recursive equation \eqref{eq:recursion} plays a central role in the definition of all particle methods. An integrated version of this equation is most often presented to emphasise the direct connection between two consecutive filtered distributions:
\begin{align}
\pi^\theta(x_n|y_{0:n}) &= \int \pi^\theta(x_{0:n}|y_{0:n}) \ud x_{0:n-1}  \label{eq:recursionfiltered} \\
&\propto g_n^\theta(y_n | x_n)\int f_n^\theta(x_n|x_{n-1}) \cdot \pi^\theta(x_{n-1}|y_{0:{n-1}}) \ud x_{n-1}.\nonumber
\end{align}

The main idea behind particle filters is to make extensive use of equation \eqref{eq:recursion} to compute sequential Monte Carlo approximations of the posterior distributions of interest, in our case, the sequence of filtered distributions. The general procedure is simple enough and mimics the iterative prediction-correction structure of any Kalman filter. A each time \(n\) the filtered density \(\pi^\theta(x_n|y_{0:n})\) can be approximated by the empirical distribution of a large sample of \(M\) (\(M>>1\)) weighted random samples termed particles. The weighted particles evolve over time: they follow the prior dynamic distribution of the model and get re-adjusted as soon as observations become available. At time \(n\), the two basic steps (a lot of refinements are possible that we will discuss later on) of particle filters are the following:
\begin{description}
 \item[Prediction:] given particles distributed along density \(\pi^\theta(x_{n-1}|y_{0:n-1})\), we simulate new particles distributed along density \(\pi^\theta(x_n|y_{0:n-1})\) with the help of the transition density \(f_{n}^\theta(x_n|x_{n-1})\).
 \item[Correction:] we re-weight these particles distributed along density \(\pi^\theta(x_n|y_{0:n-1})\) depending on the observation \(y_n\) with the help of \eqref{eq:recursion} to approximate the distribution \(\pi(x_n|y_{0:n})\).
\end{description}

Particle filters essentially combine Monte Carlo integration and importance sampling. We describe the application of self-normalised importance sampling to estimate a sequence of integrals that involve the posterior distribution \eqref{eq:dynamicposterior} and that are of the form
\begin{align*}
I_n &= \int h(x_n) \pi^\theta(x_{0:n}|y_{0:n}) \ud x_{0:n} \\
&= \int h(x_n) \pi^\theta(x_n|y_{0:n}) \ud x_n.
\end{align*}

We use the self-normalised importance sampling estimator defined in \eqref{eq:estimateur.IS2}, with \(\pi(x)=\pi^\theta(x_{0:n}|y_{0:n})\) and \(q(x)=q(x_{0:n}|y_{0:n})\). Given \(M\) particles \(X_{0:n}^1, \ldots, X_{0:n}^M\), i.i.d. with probability density \(q^\theta(x_{0:n}|y_{0:n})\), we will approximate \(I_n\) by
\begin{align*}
\widehat{I}_{n,M}^{\text{PF}} &= \sum_{j=1}^M w_n^j h(X_n^j),
\end{align*}
where mimicking the definitions given \eqref{eq:definitionpoidstilde} and \eqref{eq:definitionpoids} we define
\begin{align}
w_n^j = \frac{\widetilde{w}_n^j}{\sum_{k=1}^M \widetilde{w}_n^k},
\end{align}
with
\begin{align}
\widetilde{w}_n^j &= \dfrac{\pi^\theta(X_{0:n}^j|y_{0:n})}{q^\theta(X_{0:n}^j|y_{0:n})}.
\end{align}
Note that to alleviate the notational burden, we voluntarily omit the dependence of the importance weights on the parameter \(\theta\), and will do so for the remainder of the chapter when no confusion is possible.

\subsection*{A convenient form of importance density} Let us consider an importance density \(q\) that can be factorised as follows:
\begin{align}
q^\theta(x_{0:n}|y_{0:n})
&= q^\theta(x_n|y_{0:{n-1}}, x_{0:n})q^\theta(x_{0:{n-1}}|y_{0:{n-1}}) \nonumber\\
&= q^\theta(x_0|y_0) \prod_{k=1}^n q^\theta(x_k|y_{0:{k-1}}, y_{0:k}). \label{eq:formeduprior}
\end{align}
It is now easy to see, using \eqref{eq:recursion}, that the weights \(\widetilde{w}_n^\theta(X_{0:n}^j)\) can be updated recursively via
\begin{align}
\widetilde{w}_n^j
&= \dfrac{\pi^\theta(X_{0:n}^j|y_{0:n})}{q^\theta(X_{0:n}^j|y_{0:n})}
= \dfrac{g_n^\theta(y_n | X_n^j)f_n^\theta(X_n^j|X_{n-1}^j) \pi^\theta(X_{0:{n-1}}^j|y_{0:{n-1}})}{\pi^\theta(y_n | y_{0:{n-1}})q^\theta(X_n^j|X_{0:{n-1}}^j, y_{0:n})q^\theta(X_{0:{n-1}}^j|y_{0:{n-1}})} \nonumber \\
&= \widetilde{w}_{n-1}^j \dfrac{g_n^\theta(y_n | X_n^j)f_n^\theta(X_n^j|X_{n-1}^j)}{\pi^\theta(y_n | y_{0:{n-1}})q^\theta(X_n^j|X_{0:{n-1}}^j, y_{0:n})}. \label{eq:recursive.weights}
\end{align}
where \(\pi^\theta(y_n | y_{0:{n-1}})\) does not depend on the index \(j\), and need not be computed at all since the weights \(w_n^j\) featured in the estimator are the self-normalised version of the weights \(\widetilde{w}_n^j\) (the constant vanishes after the self-normalisation). Note that \(w_{n-1}^j\) can be substituted to \(\widetilde{w}_{n-1}^j\) in the recursive update \eqref{eq:recursive.weights} for the very same reason.

Equation \eqref{eq:recursive.weights} lies at the very core of all the particle filters in general, some variants of which we describe in the next section. It summarises, by itself, the edge that SMC methods have over MCMC methods in general in the context of dynamic models: it allows for sequential recursive estimations and predictions. At each time step, two things only are required to estimate the quantity of interest: simulations from the importance density \(q^\theta\) (the choice of which shall be discussed) and the update of the particles' weights via the computation of \eqref{eq:recursive.weights}.

\section{Particle filters}\label{sec:particlefilters}
From this point on, we adopt the convention that whenever the index \(j\) is used, we mean ``for all \(j=1,\ldots,M\)``. We present SMC methods designed to approximate the sequence of filtered distributions \(\pi^\theta(x_n|y_{0:n})\): at the end of each time step \(n\), the particle filters discussed hereafter return \(M\) particles \(X_n^j\) with weights \(w_n^j\) that can be used to approximate for instance
\begin{itemize}
 \item the filtered distribution \(\pi^\theta(x_n|y_{0:n})\) by the finite mixture of weighted Dirac masses
\begin{align*}
\widehat{\pi}(\nud x_n|y_{0:n}) &= \sum_{j=1}^M w_n^j \delta(X_n^j, \nud x_n),
\end{align*}
 \item integrals such as \(\displaystyle I_n = \int h(x_n)\pi(x_n|y_{0:n}) \ud x_n\), with \(h\in L^1(\pi(\cdot|y_{0:n}))\), by
\begin{align*}
\widehat{I}_{n,M} &= \sum_{j=1}^M w_n^j h(X_n^j).
\end{align*}
\end{itemize}

\subsection{Sequential Importance Sampling (SIS)}
\subsubsection{Conception}
The SIS filter (sometimes also called Bayesian Importance Sampling) is a direct application of the calculations shown in the previous section: it relies solely upon the sequential use of the self-normalised importance sampling technique.  The details are given in Algorithm \ref{algo:SISfilter}.
\begin{algorithm}{Sequential Importance Sampling (SIS) for filtering}\label{algo:SISfilter}~
\begin{description}
  \item[] \textbf{At time \(n=0\)}
  \begin{enumerate}
    \item Sample \(X_0^j\sim q^\theta(x_0| y_0)\).
    \item Compute \(\widetilde{w}_0^j = \dfrac{g_0^\theta(y_0 | X_0^j)\mu^\theta(X_0^j)}{q^\theta(X_0^j|y_0)}\) and set \(w_0^j \leftarrow \dfrac{\widetilde{w}_0^j}{\sum_{k=1}^M \widetilde{w}_0^k}\).
    \end{enumerate}
  \item[] \textbf{At time \(n\geq 1\)}
  \begin{enumerate}
    \item Sample \(X_{n}^j \sim q^\theta(x_n|x_{0:{n-1}}, y_{0:n})\).
    \item Compute \(\widetilde{w}_n^j = w_{n-1}^j \dfrac{g_n^\theta(y_n | X_n^j)f_n^\theta(X_n^j|X_{n-1}^j)}{q^\theta(X_n^j|X_{0:{n-1}}^j, y_{0:n})}\) and set \(w_n^j \leftarrow \dfrac{\widetilde{w}_n^j}{\sum_{k=1}^M \widetilde{w}_n^k}\).
    \end{enumerate}
    \end{description}
    \end{algorithm}
At each time step, new particles are first simulated conditionally to the old ones to represent the predictive distribution of the upcoming state and, as the observation becomes available, their weights then get readjusted to represent the filtered distribution.

\subsubsection{Prediction}
The estimation of the predicted distribution \(\pi^\theta(x_{n+\tau}|y_{0:n})\) (\(\tau\geq 1\)) can also be computed from the estimation of the filtered distribution up to time \(n\). The principle, described for instance in \cite{Doucet1998}, is identical in essence to that developed in \cite{Durbin} for Kalman filters. Since the observations at times \(n+1,\ldots,n+\tau\) are not yet available, no correction may take place after the predictions of the state that involve the transition densities \(f_{n+\tau}^\theta,\ldots,f_{n+1}^\theta\) : formally, the terms \(g_{n+\tau}^\theta, \ldots, g_{n+1}^\theta\) vanish. The details are given in Algorithm \ref{algo:SISpredict}. Observe that in this case, the importance density \(q^\theta(x_{n+\tau}|x_{0:{n+\tau-1}}, y_{0:n})\) needs to be chosen so as not to involve the yet unknown values of the upcoming observations \(y_{n+1:n+\tau}\).
\begin{algorithm}{Sequential Importance Sampling (SIS) for predicting}\label{algo:SISpredict}~
\begin{description}
  \item[]\textbf{At time \(n\geq0\), for \(\tau=1,\ldots\)}
  \begin{enumerate}
    \item Sample \(X_{n+\tau}^j \sim q^\theta(x_{n+\tau}|x_{0:{n+\tau-1}}, y_{0:n})\).
    \item Compute \(\widetilde{w}_{n+\tau}^j = w_{n+\tau-1}^j \dfrac{f_{n+\tau}^\theta(X_{n+\tau}^j|X_{n+\tau-1}^j)}{q^\theta(X_{n+\tau}^j|X_{0:{n+\tau-1}}^j, y_{0:n})}\) and set \(w_{n+\tau}^j \leftarrow \dfrac{\widetilde{w}_{n+\tau}^j}{\sum_{k=1}^M \widetilde{w}_{n+\tau}^k}\).
    \end{enumerate}
    \end{description}
    \end{algorithm}

\subsubsection{Missing observations}
When dealing with a missing observation, the SIS filter requires little modification: when observation \(Y_n\) is missing, the corresponding state \(X_n\) is predicted using Algorithm \ref{algo:SISpredict} since \(\pi^\theta(x_n|y_{0:n-1})\) is the only accessible density under such circumstances. This leads to Algorithm \ref{algo:SISmissing}.
\begin{algorithm}{Sequential Importance Sampling (SIS) for filtering with missing observations}\label{algo:SISmissing}~
\begin{description}
  \item[]\textbf{At time \(n\geq0\), if observation \(Y_n\) is missing}
  \begin{enumerate}
    \item Sample \(X_n^j \sim q^\theta(x_n|x_{0:{n-1}}, y_{0:n-1})\).
    \item Compute \(\widetilde{w}_n^j = w_{n-1}^j \dfrac{f_n^\theta(X_n^j|X_{n-1}^j)}{q^\theta(X_n^j|X_{0:{n-1}}^j, y_{0:n-1})}\) and set \(w_n^j \leftarrow \dfrac{\widetilde{w}_n^j}{\sum_{k=1}^M \widetilde{w}_n^k}\).
    \end{enumerate}
    \end{description}
    \end{algorithm}

\subsubsection{Comments}
The major drawback of the SIS filter comes from the fact that the distribution of the weights degenerates, with the variance of the importance weights increasing over time \cite[see][]{Doucet2000} meaning that the estimated distributions become less and less unreliable: after a few iterations, all but one of the normalised importance weights are close to zero. An important fraction of the calculations involved in the algorithm is thus dedicated to particles whose contributions to the estimation are almost null, making the SIS particle filter an impractical estimation procedure at best.

\subsection{Monitoring the degeneracy}\label{sec:monitordegeneracy}
To alleviate the degeneracy problem that we outlined, additional steps are traditionally implemented into Algorithm \ref{algo:SISfilter}. Since adding these new steps comes at a non negligible computational cost, it is important to somehow monitor how badly the weight distribution degenerates at a given time step, because it is usually interesting to ignore the degeneracy problem unless it reaches a given threshold.

A popular rule of thumb, first introduced in \cite{Kong1994} and later copiously reprised in the literature \cite[see for instance][]{Doucet2000,Chen,Liu2004}, is to consider the so-called effective sample size based on the normalised weights \(w_n^j\) at time step \(n\) and defined by
\begin{align*}
\frac{M}{1+\var^{q^\theta(\cdot|y_{0:n})}[w_n^1]}.
\end{align*}
This quantity is usually numerically approximated by the following estimate
\begin{align}
\mathrm{ESS}(n) &= \frac{1}{\sum_{k=1}^M (w_n^k)^2}.\label{eq:ESS}
\end{align}
It ranges from \(M\) (reached when all the particles share equal weights of value \(1\)) to \(1/M\) (reached when a single particle is given the whole probability mass of the sample, with a weight of \(1\)).

A related degeneracy measure is the coefficient of variation \cite[found in][]{Kong1994,LiuChen1995}, ranging from \(0\) to \(\sqrt{M-1}\), that is given by
\begin{align}
\mathrm{CV}(n) &= \sqrt{\frac{1}{M}\sum_{k=1}^M {(M w_n^k - 1)^2}}, \label{eq:CV}
\intertext{and satisfies to}
\mathrm{ESS}(n)&= \frac{M}{1+\mathrm{CV}(n)^2}.\label{eq:relationESSCV}
\end{align}

The Shannon entropy of the importance weights, ranging from \(\log M\) to \(0\), is sometimes also mentioned. It is defined by
\begin{align}
\mathcal{E}(n) &= -\sum_{k=1}^M w_n^k \log w_n^k. \label{eq:entropy}
\end{align}
\cite{Cornebise} recently proved that the criteria \eqref{eq:CV} and \eqref{eq:entropy} are estimators of the \(\chi^2\)-divergence and the Kullback-Leibler divergence between two distributions which are associated with the importance and target densities of the particle filter.

The evaluation of one (or more) of these criteria is introduced at each time step, with the additional procedures that we discuss next taking place if and only if the criterion reaches a certain fixed threshold so as to reduce the additional computational burden. The most common threshold found in the literature is \(\mathrm{ESS}(n) < 0.5M\). Examples illustrating the behaviours of these criteria are given later in Figures \ref{fig:essr_24_2layers_horizon5}, \ref{fig:entropy_24_2layers_horizon5} and \ref{fig:cv_24_2layers_horizon5}.

\subsection{Resample step}
A resampling step is most often introduced into Algorithm \ref{algo:SISpredict} to help and fight the degeneracy problem. The aim of this resampling step is to favour the living of the interesting particles (the ones with more important weights, that are more representative of the targeted distribution) and encourage the dying of the not so interesting particles so as to focus the computational effort upon particles that matter most for the estimation. The resampling method has to be carefully chosen, in particular it should not introduce any bias in the final estimate as mentioned in \cite{Doucet2000}

During this new step, particles are resampled according to their weights: a particle with an important weight is more likely to appear (and ''survive'') in the new sample generated, possibly more than once, whereas a particle the weight of which is close to zero is more likely not to be drawn at all (and ``die``) from a given time step to the next.

\cite{Chen} mentions that there are a few resampling schemes available in the literature. It is important to note that even though resampling might alleviate the degeneracy problem, it also brings extra random variation to the samples of particles. As a consequence, the filtered quantities of interest should preferably be computed before resampling and not after. We only present the details of the multinomial and residual resampling schemes.

\subsubsection{Multinomial resampling}
Multinomial resampling is the most popular resampling scheme, most likely because it is the easiest to both understand and implement: at a given time step, it suffices to simulate a discrete random variable which takes values \(X_n^k\) with probability \(w_n^k\). The details of multinomial resampling are given in Algorithm \ref{algo:multinomialresampling} where only the new step is described.
\begin{algorithm}{Multinomial resampling step}\label{algo:multinomialresampling}~
\begin{description}
  \item[]\textbf{At time \(n\geq 0\)}
  \begin{enumerate}
    \item[3.] Sample \(\displaystyle Z_n^j\sim \sum_{k=1}^M w_n^k\delta(X_n^k, \nud x)\).
    \item[] Replace \(X_n^j\leftarrow Z_n^j\) and \(w_n^j \leftarrow 1/M\).
    \end{enumerate}
    \end{description}
    \end{algorithm}

Used as is, it leads to the well-known Sampling Importance Resampling (SIR) filter, sometimes also called Bootstrap filter, that can be found in \cite{Gordon}. A straightforward implementation of the multinomial resampling has complexity \(\mathrm{O}(M\log M)\): it is indeed equivalent to simulating \(M\) draws from a discrete random variable \(Z_n\) such that \(\proba(Z_n=k) = w_n^k\).

A trivial implementation for such simulations requires first to draw \(U_n^1, \ldots, U_n^M\) i.i.d. with uniform distribution and then to find the indexes \(i_n^j\) for which \(U_n^j\in\interof{\sum_{k=1}^{i-1} w_n^k}{\sum_{k=1}^i w_n^k}\). Finding the indexes \(i_n^j\) has only complexity \(\mathrm{O}(M)\) when the random variables are \(U_n^j\) are ordered, but ordering these random variables has complexity \(\mathrm{O}(M\log M)\) at least, using for instance the quicksort algorithm \cite[see][]{Hoare1962}.

A practical implementation of the multinomial resampling is proposed in \cite{Doucet1998} which circumvents the naive need of sorting \(M\) i.i.d. random variables with uniform distribution and relies upon a direct simulation trick instead. The complexity of the SIR filter can hence be reduced from \(\mathrm{O}(M\log M)\) (naive implementation using quicksort) to only \(\mathrm{O}(M)\) which saves a significant amount of computational resources.

\subsubsection{Residual-multinomial resampling}
Residual-multinomial resampling is proposed in \cite{LiuChen1998} to reduce the extra variance introduced by the resamping step. It is partially deterministic as opposed to the multinomial resampling and is formulated below. Let \(\lfloor x \rfloor\) designate the integer part of a real number \(x\) and define for any \(n\geq0\):
\begin{align*}
R_n&=\sum_{k=1}^M \lfloor M\cdot w_n^k\rfloor, & \overline{w}_n^j &= \frac{M\cdot w_n^j-\lfloor M\cdot w_n^j\rfloor}{M - R_n}.
\end{align*}
\begin{algorithm}{Residual-multinomial resampling step}\label{algo:residualresampling}~
\begin{description}
  \item[] \textbf{At time \(n\geq0\)}
  \begin{enumerate}
    \item[3.] Copy \(\lfloor M\cdot\widehat{w}_n^j\rfloor\) particles \(\widehat{X}_n^j\). (\(R_n\) particles are thus allocated, say \(Z_n^1, \ldots, Z_n^{R_n}\)).
    \item[] Sample the remaining particles \(\displaystyle Z_n^{R_n+1},\ldots,Z_n^{M}\sim \sum_{k=1}^M \overline{w}_n^k\delta(X_n^k, \nud x)\).
    \item[] Replace \(X_n^j\leftarrow Z_n^j\) and \(w_n^j \leftarrow 1/M\).
    \end{enumerate}
    \end{description}
    \end{algorithm}

The details of residual-multinomial resampling are given in Algorithm \ref{algo:residualresampling} where only the new step is described. In essence, particles with weights greater than \(1/M\) are forced into the new sample, and the rest is allocated at random, depending on the remaining probability mass available. Note that the last part of a residual resampling step is basically a multinomial resampling step on the residual probability mass, hence the name.

It is shown to be computationally cheaper than the multinomial resampling, due to the fact that only a fraction of the \(M\) particles are randomly allocated. It does not introduce any bias for the estimation and has the added advantage of having a lower variance than that of the multinomial resampling \cite[see][for the proofs]{Douc2005}.

\subsubsection{Other resampling techniques}
Stratified and systematic resampling also offer an alternative to the multinomial resampling scheme (see \cite{Kitagawa1996} and \cite{Carpenter1999} or \cite{Chen} for a more general overviews). Systematic resampling appears to be another popular choice in the literature for computational reasons even though its variance is not guaranteed to be smaller than that of the multinomial resampling as stated in \cite{Douc2005}. A short study of these techniques and a numerical comparison of their performance on an example are offered in \cite{Cornebise}. Note that residual versions of these techniques also exist, where they are substituted to the multinomial sampling used in the second half of Algorithm \ref{algo:residualresampling}.

\subsubsection{Limitations of the resampling procedure}
The resampling procedure alleviates the degeneracy problem but also introduces practical and theoretical issues \cite[as mentioned in][for example]{Doucet2000}. From a practical point of view, resampling very obviously limits the opportunity of parallelisation of the algorithm. From a theoretical point of view, simple convergence results are lost due to the fact that after one resampling step the particles are not independent anymore. Moreover, resampling causes the particles with high importance weights to be statistically selected many times: the algorithm thus suffers from the so-called loss of diversity.

\subsection{Move step}
The loss of diversity among the particles following the resample step is usually addressed in the literature with the introduction of yet another additional move step into the algorithm: the idea behind it is to rejuvenate the diversity after the particles have been resampled.
\subsubsection{Using MCMC}
\cite{Gilks,DoucetDeFreitas} present the so-called Resample-Move algorithm in which an MCMC step is used after resampling. This new step relies upon the use of Markov transition kernels with appropriate invariant distributions. Moving the particles according to such kernels formally guarantees the particles still target the distribution of interest but also give them an additional chance to move towards an interesting region of the state space while increasing the diversity of the sample at the cost of an increased computational burden. \cite{DoucetJohansen} underline the possibility of using even non ergodic MCMC kernels for this purpose and also propose to go a step further and rejuvenate not only the current state but also some of the (immediate) past states with the so-called Block Sampling (the computational cost of which is thus even greater).

\subsubsection{Using regularisation}
Another approach to deal with the loss of diversity is based upon regularisation techniques. Let us define for \(x, x^*\in\mathcal{X}\subset\R^{n_x}\)
\begin{align*}
K_h(x, x^*) &= h^{-n_x} \cdot (\det \Sigma_n)^{-1/2} \cdot K\left(\Sigma_n^{-1/2}\cdot\frac{x-x^*}{h}\right)
\end{align*}
where \(K\) is usually a smooth symmetric unimodal positive kernel of unit mass (hence a probability measure), \(h\) is the bandwidth of the kernel, and \(\Sigma_n\) designates the empirical covariance matrix of the sample \cite[see][for the idea of whitening the sample via \(\Sigma_n\)]{Silverman}.

\begin{algorithm}{Regularisation step}\label{algo:regularisation}~
\begin{description}
  \item[] \textbf{At time \(n\geq 0\)}
  \begin{enumerate}
    \item[4.] Sample \(\epsilon_n^j\sim K(x)\), and set \(Z_n^j \leftarrow X_n^j + h\cdot\Sigma_n^{1/2}\cdot\epsilon_n^j\).
    \item[] Replace \(X_n^j\leftarrow Z_n^j\) and keep \(w_n^j \leftarrow w_n^j\).
    \end{enumerate}
    \end{description}
    \end{algorithm}
\cite{Gordon} originally referred to that step as ''jittering`` since it adds a small amount of noise to each resampled particle. Note that, when used together with the multinomial resampling scheme described in Algorithm \ref{algo:multinomialresampling}, the resulting combination of the two steps can be reformulated as described in Algorithm \ref{algo:multinomialregularisation}: it is then equivalent to resampling new particles from the smoothed estimated target distribution (using kernel density \(K\)).
\begin{algorithm}{Alternate formulation for the combination of Algorithms \ref{algo:multinomialresampling} and \ref{algo:regularisation}}\label{algo:multinomialregularisation}~
\begin{description}
  \item[] \textbf{At time \(n\geq 0\)}
  \begin{enumerate}
    \item[3+4.] Sample \(\displaystyle Z_n^j\sim \sum_{k=1}^M w_n^k K_h(X_n^k, x)\).
    \item[] Replace \(X_n^j\leftarrow Z_n^j\) and \(w_n^j \leftarrow 1/M\).
    \end{enumerate}
    \end{description}
    \end{algorithm}

The choice of both the kernel smoothing density \(K\) and the bandwidth \(h\) obviously has a big impact on the algorithm. The idea is to resample from a density estimated from the particles at time step \(n\) that best approximates the true target density. Picking \(K(\cdot)=\delta(\cdot, 0)\) the Dirac mass at the origin turns the regularised SMC filter back into a simple SMC filter. From a general point of view, we would like the estimated density to converge as fast as possible towards the true target density as \(M\) goes to \(+\infty\), since the number of particles will necessarily be limited by the computational resources.

For the Gaussian kernel (among others), \cite{Silverman} shows it is possible to compute the optimal bandwidth to use, i.e. the bandwidth that minimises the variance of the density estimate. Although it could be argued that selecting a proper bandwidth is a difficult task, this optimal bandwidth yields good results in practise and at least provides a rough idea about the scaling of \(h\). As is the case with kernel density estimates, the choice of \(h\) directly influences the trade-off made between variance and bias of the estimate: if \(h\) is chosen too small, the loss of diversity will still be severe, and if \(h\) is chosen too large, the filtered density will roughly be estimated as a single kernel, hence introducing a severe bias into the estimation.

The use of the Epanechnikov kernel, proportional to \(1-\|x\|^2\) on the unit ball of the state space, is recommended in \cite{Silverman} because it is asymptotically the most efficient, and \cite{Doucet1998} claims it can be difficult to choose a ''good'' kernel. However, we advocate the use of the Gaussian kernel whenever possible for computational reasons: simulations from the Gaussian kernel are readily available on most machines and come at a computationally cheaper price than simulations from the Epanechnikov kernel. But the non optimality of Gaussian kernel does not outbalance its ease of use, since the choice of the kernel neither affects the order of the bandwidth nor the rate of convergence as stated in \cite{DasGupta}.

From a general point of view it is also possible to choose a \(n_x\)-dimensional kernel under the form of a product of \(n_x\) 1-dimensional (possibly distinct) kernels. Such a choice is preferable when some coordinates of the state are bounded. It allows for easier simulations on these coordinates using dedicated truncated kernels whereas a straightforward accept-reject algorithm could turn out to be highly inefficient (with a low acceptance rate) depending on the boundaries of the state space.

Finally, the regularisation can also be done before resampling thus resulting in the so-called pre-regularised particle filter (pre-PRF) as opposed to the post-regularised particle filter presented here. Theoretical convergence results about these regularised filters are available in \cite{Oudjane} and \cite{Rossi}

\subsection{Detection and removal of outliers}
In order to deal with the sensitivity of the particle filters to outliers, we propose a new additional rule at the end of step 2 of Algorithm \ref{algo:SISfilter}. Its role is to make sure that outliers do not lead to a fully degenerated situation, that the algorithm would not recover from. The details of it are given in Algorithm \ref{algo:ignoreoutliers} where only the additional rule is described.
\begin{algorithm}{Online detection and removal of outliers.}\label{algo:ignoreoutliers}~
\begin{description}
  \item[] \textbf{At time \(n\geq 0\)}
  \begin{enumerate}
    \item[] If the degeneracy problem is critical, consider the observation \(y_n\) as missing (see Algorithm \ref{algo:SISmissing}) and rewind back to step 1.
    \end{enumerate}
    \end{description}
    \end{algorithm}
The rule applies only to situations where the degeneracy of the sample is critical: when the importance density chosen is the prior density, it triggers only when the current observation is not predicted efficiently. In that case, we proceed as if the observation was missing. In practise the degeneracy problem is deemed critical when a criterion such as \(\mathrm{ESS}(n)<\epsilon\cdot M\) is met, with \(\epsilon>0\) very small.

Observations that do not agree with the model are thus detected online and ignored to prevent immediate degeneracy. This trick is in a way similar to the one introduced in \cite{HuSL2008,HuSL2011} where a resample step is iterated until the likelihood of the current observation with regard to the resampled particles is above a given threshold. While both techniques ensure that the particles do not collapse when an outlier is met, the cost paid is different for each. The alteration proposed in \cite{HuSL2008,HuSL2011} can be computationally expensive (with an unbounded runtime) but the observation ends up being taken into account, while our own modification is definitively cheaper (with a guaranteed fixed runtime) but discards the observation at hand when it strongly disagrees with the current state of the model. A significant change of state will still be detected in the long run, because considering the observation \(y_n\) as missing automatically implies the variance of the state grows larger (which means that, if it were to be repeated, the outlying observation, would seem more likely at the next time step, with regard to the new state).

\subsection{Choice of the importance density}\label{subsec:importancedensity}
As previously stated the particle filters rely on the introduction of an importance density that was chosen of the form given in \eqref{eq:formeduprior} i.e.
\begin{align*}
q^\theta(x_{0:n}|y_{0:n})
&= q^\theta(x_0|y_0) \prod_{k=1}^n q^\theta(x_k|y_{0:{k-1}}, y_{0:k}).
\end{align*}
Choosing carefully the importance density \(q^\theta\) can help reduce the variance of the importance weights and thus alleviate the degeneracy problem. As the choice is abundantly discussed in the literature, we only selected three representative alternative among the many that are available.
\subsubsection{Prior density}
A default choice consists of taking \(q^\theta(x_0|y_0) = \mu^\theta(x_0)\) and \(q^\theta(x_n|x_{0:{n-1}}, y_{0:n}) = f_n^\theta(x_n|x_{n-1})\), i.e. taking the prior density \eqref{eq:dynamicprior} of the model as the importance function. This choice works even with missing data (as it does not depend on \(y_n\)) and leads to much simpler calculations for the update of the importance weights as can be seen directly in the formulae given in Algorithm \ref{algo:SISfilterwithpriorasimportance}.
\begin{algorithm}{Sequential Importance Sampling (SIS) for filtering, using the prior density as the importance density}\label{algo:SISfilterwithpriorasimportance}~
\begin{description}
  \item[] \textbf{At time \(n=0\)}
  \begin{enumerate}
    \item Sample \(X_0^j\sim \mu^\theta(x_0)\).
    \item Compute \(\widetilde{w}_0^j = g_0^\theta(y_0 | X_0^j)\) and set \(w_0^j \leftarrow \dfrac{\widetilde{w}_0^j}{\sum_{k=1}^M \widetilde{w}_0^k}\).
    \end{enumerate}
  \item[] \textbf{At time \(n\geq 1\)}
  \begin{enumerate}
    \item Sample \(X_{n}^j \sim f_n^\theta(x_n|X_{n-1}^j)\).
    \item Compute \(\widetilde{w}_n^j = w_{n-1}^j g_n^\theta(y_n | X_n^j)\) and set \(w_n^j \leftarrow \dfrac{\widetilde{w}_n^j}{\sum_{k=1}^M \widetilde{w}_n^k}\).
    \end{enumerate}
    \end{description}
    \end{algorithm}
Note that using the prior density as the importance density makes the algorithm propose new particles in a blind way: the new particles are simulated around the current state, not around the upcoming targeted state. With such a choice of importance density, the algorithm becomes especially sensitive to outliers. An annealed version of the prior distribution is proposed in \cite{Chen} to help deal with some situations where prior and likelihood do not agree.

\subsubsection{Optimal density}
Although popular, the choice of the prior density is not optimal: the optimal choice is given by \(q^\theta(x_0|y_0) = \pi^\theta(x_1|y_1)\) and \(q^\theta(x_n|x_{0:{n-1}}, y_{0:n}) = \pi^\theta(x_n|y_n,x_{n-1})\) in the sense that it minimises the variance of the importance weights conditional upon the past states and the past observations as can be seen in \cite{Doucet2000}. The idea underlying this choice is to take into account the upcoming observation so that particles are not blind to the upcoming state anymore. Most of the time sampling from these optimal distributions is not an option however, and it is usually recommended to approximate them if possible. For example \cite{PittShephard} propose the so-called Auxiliary Particle Filter (APF) which essentially reverses the sampling and resampling phase mentioned in the previous algorithms \cite[see][]{WhiteleyJohansen2011}. It relies upon the introduction of an augmented state that is used to select the most representative particles in the sense that their predictive likelihoods are large. \cite{Doucet2000} use the Extended Kalman filter to derive a Gaussian approximation (relying on a local linearisation of the state space model) and \cite{vanderMerwe} discuss the use of the Unscented Kalman Filter to obtain such approximations \cite[see][for the details about implementing the UKF]{Wan2000Unscented}.

\subsubsection{Independent density}
Let us mention that it is also possible to use an independent importance density (independent with regard to the states and observations) but it is strongly recommended to avoid such a choice because it ''ignores'' both the current and the upcoming states \cite[see][]{Doucet2000}

\subsection{Parameter estimation}\label{subsec:parameterestimation}
Thus far, state estimation was discussed conditionally to the fact that the parameter \(\theta\) was known. However, \(\theta\) is often unknown and has to be estimated together with the state of the dynamic model. \cite{Kantas} offers a comparative review of the possible choices available for parameter estimation, presenting maximum likelihood and Bayesian parameter estimation in the context of an offline or online procedure. We provide here only a brief overview of the Bayesian parameter estimation and direct the interested reader to the original paper for the complete discussion.

One of the first approach considered in the literature for parameter estimation is to extend the state \(X_n\) at time \(n\) into a new state \(Z_n=(X_n,\theta_n)\) with initial distribution \(\mu^{\theta_0}(x_0)\pi(\theta_0)\) and transition density \(f^{\theta_n}(x_n|x_{n-1})\cdot\delta(\theta_n,\theta_{n-1})\) and then estimate this new extended model with a standard SMC filter as in \cite{Kitagawa1996}. Even though the approach is theoretically sound as claimed in \cite{Rossi,Kantas}, it can lead to a strong loss of diversity problem on the coordinate \(\theta\) when no move step is implemented as the parameter space is only explored at the initialisation of the algorithm, making such an approach often unusable.

The addition of a move step into the algorithm provides a satisfying solution to this problem as can be seen in \cite{Rossi} who successfully applied the kernel regularisation technique, or in \cite{Andrieu1999} who makes use of MCMC techniques in a move step to update the parameter value. The regularisation can also be combined with a judicious choice of importance density such as with the APF \cite[see][]{PittShephard} to provide remarkably accurate parameter estimation \cite[as shown in][for example]{CasarinMarin2009,WhiteleyJohansen2011}. Another option is to force a fictitious small dynamic upon the parameter as described in \cite{Kitagawa1998,Higuchi01} so that it is artificially allowed to evolve over time, even though \cite{Kantas} rightly remarks that modifying the model in such a way makes it hard to quantify how much bias is introduced in the resulting estimates.

A more recent way of estimating the parameter together with the state relies upon the use of so-called Particle Markov Chain Monte Carlo (PMCMC) methods found in \cite{AndrieuDoucetHolenstein}. These methods are computationally expensive both in term of storage and calculations, because their computational cost typically grows with time as underlined in \cite{ChopinJacob2012}, and thus are less fit for online estimation than some standard SMC filter: the most basic PMCMC method, known as the Particle Marginal Metropolis-Hastings (PMMH) sampler and described in \cite{Kantas}, involves running an SMC filter for each step of a Metropolis-Hastings algorithm used to propose a new value of the parameter \(\theta\).

\subsection{Summary}\label{sec:particlefilterused}
In the end, keeping in mind that the original aim is the online estimation and prediction, we implemented an algorithm not too computationally expensive. We chose the importance density to be the prior density of the model and included a residual resample step coupled with a regularisation move step, that triggered whenever \(\mathrm{ESS}(n) < 0.5M\) unless \(\mathrm{ESS}(n) < 0.001M\), in which case the current observation was instead considered an outlier and thus treated as missing. For the regularisation step (see Algorithm \ref{algo:regularisation}) we use a Gaussian kernel \(K\) with a bandwidth \(h\) optimally chosen for the mean integrated squared error \cite[see][Chapter 4]{Silverman}. As for the parameter estimation problem, we opted for the solution of extending the state-space and introduced no artificial dynamic on the parameter \(\theta\) (thus using \(\theta_{n}=\theta_{n-1}\)). In practise, this results in the disappearance of the \(\theta\) superscript on densities \(\mu\), \(f_n\) and \(g_n\) in the description of Algorithm \ref{algo:filterfinal}. We did however test the introduction of an artificial dynamic on the parameters but observed no changes in the measured overall performance. Note that the regularisation step mentioned above applies to the extended state (i.e. including the parameter). 

\begin{algorithm}{Particle filter used for our application}\label{algo:filterfinal}~
\begin{description}
  \item[] \textbf{At time \(n=0\)}
  \begin{enumerate}
    \item Sample \(\widehat{X}_0^j\sim \mu(x_0)\).
    \item Compute \(\widetilde{w}_0^j = g_0(y_0 | X_0^j)\) and set \(\widehat{w}_0^j \leftarrow \dfrac{\widetilde{w}_0^j}{\sum_{k=1}^M \widetilde{w}_0^k}\).
    \begin{itemize}
      \item if \(\widehat{\mathrm{ESS}}(0) < 0.001M\), set \(X_0^j \leftarrow \widehat{X}_0^j\) and \(w_0^j \leftarrow 1/M\).
      \item if \(0.001M \leq \widehat{\mathrm{ESS}}(0) < 0.5M\), use residual-multinomial resample (see Algorithm \ref{algo:residualresampling}) and regularisation move (see Algorithm \ref{algo:regularisation}) steps to set \(X_0^j\) and \(w_0^j\).
      \item if \(0.5M \leq \widehat{\mathrm{ESS}}(0)\), set \(X_0^j \leftarrow \widehat{X}_0^j\) and \(w_0^j \leftarrow \widehat{w}_{0}^j\).
      \end{itemize}
    \end{enumerate}
  \item[]\textbf{At time \(n\geq 1\)}
  \begin{enumerate}
    \item Sample \(\widehat{X}_{n}^j \sim f_n(x_n|X_{n-1}^j)\).
    \item Compute \(\widetilde{w}_n^j = w_{n-1}^j g_n(y_n | X_n^j)\) and set \(\widehat{w}_n^j \leftarrow \dfrac{\widetilde{w}_n^j}{\sum_{k=1}^M \widetilde{w}_n^k}\).
    \begin{itemize}
      \item if \(\widehat{\mathrm{ESS}}(n) < 0.001M\), set \(X_n^j \leftarrow \widehat{X}_n^j\) and \(w_n^j \leftarrow w_{n-1}^j\).
      \item if \(0.001M \leq \widehat{\mathrm{ESS}}(n) < 0.5M\), use residual-multinomial resample (see Algorithm \ref{algo:residualresampling}) and regularisation move (see Algorithm \ref{algo:regularisation}) steps to set \(X_n^j\) and \(w_n^j\).
      \item if \(0.5M \leq \widehat{\mathrm{ESS}}(n)\), set \(X_n^j \leftarrow \widehat{X}_n^j\) and \(w_n^j \leftarrow \widehat{w}_{n}^j\).
      \end{itemize}
    \end{enumerate}
    \end{description}
    \end{algorithm}

\section{Application}\label{sec:applicationdynamic}
In this Section we describe an application of particle filters for electricity load forecasting. We quickly describe the data used for our experimentation and the two similar models that were estimated using Algorithm \ref{algo:filterfinal}, deal with the problem of initialising the particle filter and discuss the results obtained.
\subsection{Data}
\subsubsection{Time range}
The data chosen for the application contain the consolidated half-hourly electricity load at the ''EDF'' perimeter over the period ranging from 04/01/2006 to 03/31/2011 which represents five years worth of measurements, with 48 points per day. Note that only an estimation of the load is available in real time. The consolidated data correspond to the true (not estimated) signal that is available only three weeks later.
\subsubsection{Daytypes}
The calendar used for the application provides nine distinct daytypes, the list of which is given in Table \ref{tab:cal_simple}. In essence, this is a very basic calendar that models a single bank-holidays effect where more detailed calendars would model multiple different ones. Although such a basic calendar arguably does not reflect the whole variety of daytypes, it is detailed enough for our purpose and helps keep the dimension of the model we propose as low as possible.
\begin{table}[htbp]
\begin{center}
\begin{tabular}{cc}
  \hline
 \# & day \\
  \hline
 0 & mon.  \\
 1 & tue.-wed.-thu. \\
 2 & fri. \\
   \hline
\end{tabular}\hspace{3em}
\begin{tabular}{cc}
  \hline
 \# & day \\
  \hline
 3 & sat. \\
 4 & sun. \\
 5 & before BH \\
 \hline
\end{tabular}\hspace{3em}
\begin{tabular}{cc}
  \hline
 \# & day \\
  \hline
 6 & BH \\
 7 & after BH \\
 8 & between BH and a weekend \\
 \hline
\end{tabular}
\end{center}
\caption{Daytypes provided by the basic calendar used in the application. BH stands for a bank holiday.}
\label{tab:cal_simple}
\end{table}

Note that the operational model used by EDF also require the precise specification of daytypes and so-called offsets, the latter being used to model breakpoints \cite[see][for the details]{Bruhns}.

\begin{figure}[htbp]
\begin{center}
\includegraphics[width=1\textwidth]{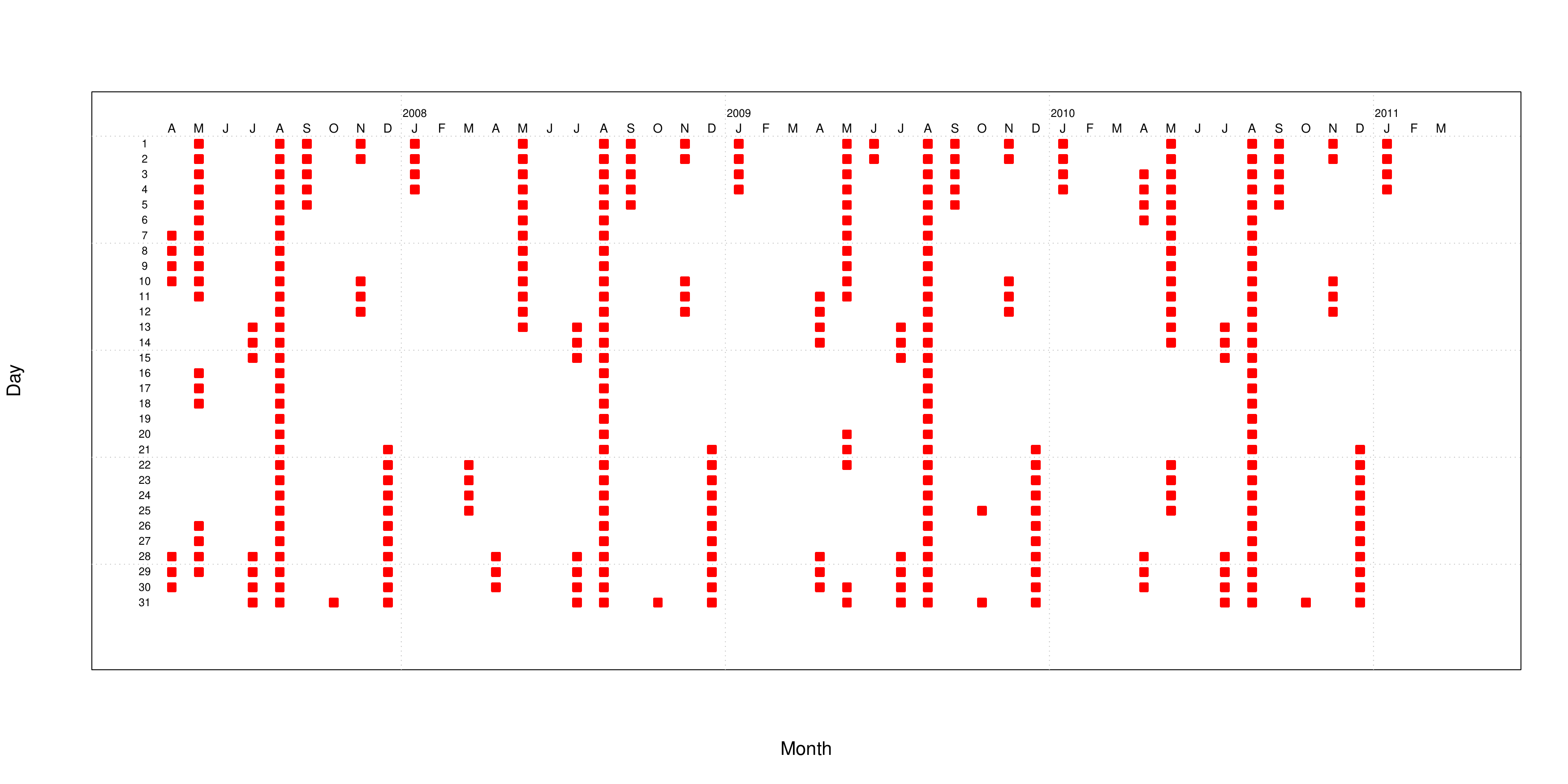}
\end{center}
\caption{Repartition of the bank holidays amidst the calendar from 04/01/2007 to 03/31/2011. Each column represents a month.}
\label{fig:calendar_dayvalidity}
\end{figure}

From here on, we will call bank-holidays, the instants in the calendar where specific information is needed for the operational model to be correctly estimated and predicted. These instants essentially correspond to bank-holidays (daytypes from \(5\) to \(8\), hence the name), or the summer and winter holiday breaks and are signalled on Figure \ref{fig:calendar_dayvalidity}.

\subsection{Dynamic model}\label{sec:dynamicmodels}
The formulation of the model that we consider was inspired by the works of \cite{Bruhns,Dordonnat2,Launay1}. It features three parts (seasonality, heating, cooling) similarly to what was done in \cite{Launay1} and includes a two layers dynamic on the two most relevant parts with regard to the French electricity load for each of the 48 half-hours (or instants) within a day. The 48 corresponding independent model is estimated and predicted in parallel, using the calendar and temperature information described above, the results being aggregated back together at the end of the process. The dimensions of the parameter and state spaces were voluntarily kept small: the goal is ultimately to provide competitive one-day-ahead predictions for the electricity load based on a model as parsimonious as possible within a rather general framework.

We denote \(\loi{N}(\mu, \Sigma)\) the Gaussian distribution with mean \(\mu\) and variance \(\Sigma\), and \(\loi{N}(\mu, \Sigma, \mathcal{S})\) the corresponding truncated Gaussian distribution the support of which is \(\mathcal{S}\). For each half-hour, and removing the now superfluous \(i\) subscript, the model that we consider is defined by
\begin{align}
y_n &= x_n+\nu_n, \label{eq:maindynamicmodel}
\end{align}
where \(\nu_n \sim \loi{N}(0, \sigma^2)\) and where the state \(x_n\) is made of three parts
\begin{align*}
x_n &= x_n^{\text{season}} + x_n^{\text{heat}} + x_n^{\text{cool}},
\end{align*}
that are defined by
\begin{align*}
x_n^{\text{season}} &= s_n \cdot \kappa_{\text{daytype}_n} \\
x_n^{\text{heat}} &= g_n^{\text{heat}} (T_n^{\text{heat}} - u^{\text{heat}})\one_{\interoo{T_n^{\text{heat}}}{+\infty}}(u^{\text{heat}}) \\
x_n^{\text{cool}} &= g^{\text{cool}} \Delta_n^{\text{cool}}.
\end{align*}
The various components obey the following prior dynamic
\begin{align*}
s_n &= s_{n-1} + \epsilon_n^s, & \epsilon_n^s &\sim \loi{N}\left(0, \sigma_{s,n}^2, \interoo{-s_{n-1}}{+\infty}\right) \\
g_n^{\text{heat}} &= g_{n-1}^{\text{heat}} + \epsilon_n^g, & \epsilon_n^g &\sim \loi{N}\left(0, \sigma_{g,n}^2, \interoo{-\infty}{-g_{n-1}^{\text{heat}}}\right) \\
\sigma_{s,n} &= \sigma_{s,n-1} + \eta_n^s, & \eta_n^s &\sim \loi{N}\left(0, \sigma_s^2, \interoo{-\sigma_{s,n-1}}{+\infty}\right) \\
\sigma_{g,n} &= \sigma_{g,n-1} + \eta_n^g, & \eta_n^g &\sim \loi{N}\left(0, \sigma_g^2, \interoo{-\sigma_{g,n-1}}{+\infty}\right)
\end{align*}
where \(\text{daytype}_n\), \(T_n^{\text{heat}}\) and \(\Delta_n^{\text{cool}}\) correspond to the exogenous variables that we already discussed:
\begin{itemize}
\item denoting \(N_{\text{daytype}}\) the number of different daytypes featured in the calendar provided, \(\text{daytype}_n \in \N\) takes a finite number of values between \(0\) and \(N_{\text{daytype}}-1\) and represents the class to which the day \(n\) belongs with regard to the calendar ;
\item \(T_n^{\text{heat}} \in \R\) is the temperature used to compute the heating part of the model, which is precomputed at EDF as a mixture of exponentially smoothed signals ;
\item \(\Delta_n^{\text{cool}} \in \R_+\) provides the cooling degrees needed to compute the cooling part of the model.
\end{itemize}

Using the definitions and notations introduced in Section \ref{sec:intro3}, the parameter of the model is given by \(\theta = (\sigma_s, \sigma_g, g^{\text{cool}}, u^{\text{heat}}, \kappa, \sigma)\), these quantities are assumed constant over time in the model. At time \(n\), the state of the model is given by \(x_n\) whose components \((s_n, g_n, \sigma_{s,n}, \sigma_{g,n})\in\R^4\) are the quantities that vary over time according to the dynamic specified. All these quantities are unknown and are to be estimated.

The model \eqref{eq:maindynamicmodel} includes a seasonal part \(x_n^{\text{season}}\) that is essentially made of a signal \(s_n\), the dynamic prior of which is a random-walk process whose standard deviation \(\sigma_{s,n}\) itself evolves as a random-walk. \(s_n\) is multiplied by a coefficient \(\kappa_{\text{daytype}_n}\) that depends on the daytype of the current observation to model the difference in behaviour between the electricity load on weekdays and weekends or holidays. For identifiability reason, the sum of the coefficients \(\kappa_j\) is fixed so that
\begin{align*}
\dfrac{1}{N_{\text{daytype}}}\sum_{j=1}^{N_{\text{daytype}}} \kappa_j = 1.
\end{align*}
Note that \(s_n\) essentially replaces the truncated Fourier series featured in \cite{Launay1}.

The model \eqref{eq:maindynamicmodel} also includes two weather-related parts to account for the influence of low (heating part) and high temperatures (cooling part) upon the electricity load : the heating part \(x_n^{\text{heat}}\) is based upon a truncated difference between the temperature \(T_n^{\text{heat}}\) and a heating threshold \(u^{\text{heat}}\), as studied in \cite{Launay2}. This difference is multiplied by a gradient \(g_n^{\text{heat}}\) whose dynamic is similar to that of \(s_n\): the prior is a random-walk whose standard deviation \(\sigma_{g,n}\) itself evolves as a random-walk. Because the cooling effect in France is of a lesser magnitude than the heating effect, the corresponding model for the cooling part \(x_n^{\text{cool}}\) is simpler: the precomputed truncated difference \(\Delta_n^{\text{cool}}\) is given to the model and multiplied by a cooling gradient \(g^{\text{cool}}\).

Notice that to ensure the different quantities involved kept consistent signs throughout time, we specifically used truncated Gaussian distributions. In particular, this means that the random-walks featured in the dynamic are not symmetric and hence that the mean of the state is a priori expected to slightly evolve over time. The constraint on \(\epsilon_n^s\) and \(\epsilon_n^g\) can of course easily be lifted if need be, and does not affect the overall predictive performance of the model in any way.

\subsection{Initialisation of the particle filter}\label{subsec:initialisation}
As was already discussed, the degeneracy of the particles sample over time is a serious matter. The choice of the initial distribution of the state is thus of the utmost importance because a strong disagreement between this distribution and the first filtered distribution could lead to sample degeneracy after only a single time step. Two solutions are theoretically viable to choose the initial prior distribution: one may choose either a vague or an informative distribution.

\begin{enumerate}
  \item on one hand, a vague prior has the advantage of not biasing the dynamic model before the first observations. However, the variance of the initial distribution of the state being very large, the sequence of posterior variances of the filtered distributions of the state tend to decrease very quickly at first. From an SMC filter point of view, one has to use a very large sample of particles to cover at the same time the regions of the state space with prior highest probability and with posterior highest probability: a vague initialisation thus requires the use of a massive number of particles.

  \item on the other hand, designing an informative distribution is a totally different task, but still not a trivial one: one has to keep in mind that a ``bad`` choice of initial distribution may lead to immediate degeneracy. Intuitively, the ideal solution would be to dispose at time \(n=-1\) of a filtered distribution \(\pi^\theta(x_{-1}|y_{-1}, \ldots, y_{-m})\) to use it as a the initial distribution at time \(n=0\). Such a choice is of course not possible because observations are only available for time \(n=0,\ldots,N\).
  \end{enumerate}
Note that the trick of ignoring outliers introduced into the particle filter (see Algorithm \ref{algo:ignoreoutliers}) does not alleviate the problem of initialisation, since it can only increase the variance if it is used.

We thus opted for a more general procedure that allows for an automated initialisation of the particles sample to a fitting state space region from time \(n=n_0\), and that combines the two approaches mentioned above to retrieve the benefits of both:
\begin{enumerate}
  \item we use a vague distribution to estimate the smoothed distribution up to time \(n=n_0-1\) using open-source MCMC generic software such as BUGS \citep{WinBUGS} or JAGS \citep{JAGS}: we typically chose \(n_0=365\) so that the variance of the filtered distribution at time \(n=n_0-1\) is already small enough not to require the use of a massive amount of particles ;
  \item after this first MCMC initialisation phase, we retrieve particles (approximately) distributed along the filtered distribution of the state at time \(n=n_0-1\): this distribution is the one used (through these particles) to initialise the SMC filter at time \(n_0\).
\end{enumerate}
There is however a price to pay for solving the initialisation problem in such a way. First we have to use MCMC to initialise the particle filter and second it makes it hard to use the particle filter on a time series with few observations. Note that the first issue raised is but rhetorical: MCMC, even if expensive, has to be run only once, and not at each time step.

\subsubsection{Initial distribution for the MCMC estimation}
The initial distribution envisioned for the dynamic model \eqref{eq:maindynamicmodel} is vague and specified by:
\begin{align*}
s_0, g^{\text{cool}}   &\sim \loi{N}(0, 10^8, \R_+) \\
g_0^{\text{heat}}   &\sim \loi{N}(0, 10^8, \R_-)   \\
u^{\text{heat}}     &\sim \loi{N}(14, 1)      \\
\kappa/N_{\text{daytype}}  &\sim \loi{D}_{N_{\text{daytype}}}(1,\ldots,1) \\
\sigma^2, \sigma_{s,0}^2, \sigma_{g,0}^2, \sigma_{s}^2, \sigma_{g}^2 &\sim \loi{IG}(10^{-2},10^{-2})
\end{align*}
where \(\loi{D}_{d}(\alpha_1,\ldots,\alpha_d)\) is the Dirichlet distribution in \(\R_+^d\) with parameter \(\alpha\) (in particular \(\loi{D}_{d}(1,\ldots,1)\) is the uniform distribution over the simplex of \(\R_+^d\) defined by \(\sum_{i=1}^d x_i = 1\)).

\subsubsection{Practical issue}
We faced some technical issues running the MCMC estimation up until time \(n_0=365\) since the Markov Chain outputs were not usable: even with a large burn-in period, the sample returned would not pass the diagnostic tests for the convergence of the empirical distribution towards the true target \cite[see][for example]{GelmanRubin}. For the initialisation via MCMC we thus separated the initial distribution into two parts, essentially isolating the dynamic on the variance of the random-walks, and proceeded as follows.

First we estimated the model as defined in \eqref{eq:maindynamicmodel} up until time \(n_0=365\), using MCMC generic software such as described in \cite{WinBUGS, JAGS}, with the following modification
\begin{align*}
\sigma_{s,n} &= \sigma_{s,n-1} = \sigma_{s,*} \\
\sigma_{g,n} &= \sigma_{g,n-1} = \sigma_{g,*}
\intertext{with initialisation}
\sigma_{s,*}^2,\sigma_{g,*}^2 &\sim \loi{IG}(10^{-2},10^{-2}),
\end{align*}
in essence removing the second layer in the dynamic from the model (since \(\sigma_{s,n}\) and \(\sigma_{s,n}\) are not allowed to vary with time anymore). This led to a posterior distribution on a diminished state, that we denote \(\widetilde{\pi}_1(\widetilde{x}_{n_0-1}|y_{0:n_0-1})\). From there we completed this posterior distribution with an additional prior \(\widetilde{\pi}_2\) on \(\sigma_{s}\) and \(\sigma_{g}\) to serve as an initialisation at for the full model at time \(n_0\).

The initial distribution of the particle filter for the full model at time \(n_0\) was thus of the form
\begin{align*}
\pi(x_{n_0-1}|y_{0:n_0-1}) &\propto \widetilde{\pi}_1(\widetilde{x}_{n_0-1}|y_{0:n_0-1}) \times \widetilde{\pi}_2(\sigma_{s,n_0-1},\sigma_{g,n_0-1})
\end{align*}
with
\begin{align*}
\sigma_{s}^2 &\sim \loi{N}(\overline{m}_s, \overline{s}_s^2,\R_+^*)   \\
\sigma_{g}^2 &\sim \loi{N}(\overline{m}_g, \overline{s}_g^2,\R_+^*)
\end{align*}
where \(\overline{m}_s, \overline{m}_g, \overline{s}_s^2,
\overline{s}_g^2\) were values chosen empirically based on
\(\widetilde{\pi}_1\).\\  For example, we chose \(m_s\) and \(m_g\) to
be the standard deviations of the posterior MCMC estimated samples  \((\esp[\epsilon_1^s|y_{0:n_0}],\ldots,\esp[\epsilon_{n_0}^s|y_{0:n_0}])\) and \((\esp[\epsilon_1^g|y_{0:n_0}],\ldots,\esp[\epsilon_{n_0}^g|y_{0:n_0}])\) respectively.

\subsection{Predictions}
\subsubsection{Quality criterion}
To assess the quality of the models we propose, we will mainly look at their respective predictive performances measured by Mean Absolute Percentage Error (MAPE). As a matter of fact, we are working with half-hourly data and we will model each half-hour independently from one another, a common choice given the type of data, thus leading to 48 separate daily model (see Section \ref{sec:dynamicmodels}). Indexing the respective MAPE criteria of these models by the instant \(i=0,\ldots,47\) to which they are associated, and given their respective observations \(y_{1,i},\ldots,y_{n,i}\), these models return 48 \(\tau\)-day-ahead predictions defined as the expectations of the predictive distributions i.e. for \(i=0,\ldots,47\)
\begin{align}
\widehat{y}_{n+\tau,i} &= \esp[x_{n+\tau}|y_{0:n,i}]\label{eq:prediction}.
\end{align}
The corresponding predictive (with prediction horizon \(\tau\)) MAPE criterion that we consider for these 48 models is defined, for \(i=0,\ldots,47\), by
\begin{align*}
\mathrm{MAPE}_i(\tau) &= \frac{1}{n-\tau}\sum_{k=1}^{n-\tau} \left|{\frac{\widehat{y}_{k+\tau,i} - y_{k+\tau,i}}{y_{k+\tau,i}}}\right|
\end{align*}
and we will most often aggregate the results as
\begin{align*}
\mathrm{MAPE}(\tau) &= \frac{1}{48}\sum_{i=0}^{47} \mathrm{MAPE}_i(\tau) \\
&= \frac{1}{48(n-\tau)}\sum_{i=0}^{47}\sum_{k=1}^{n-\tau} \left|{\frac{\widehat{y}_{k+\tau,i} - y_{k+\tau,i}}{y_{k+\tau,i}}}\right|.
\end{align*}

\subsubsection{Operational predictions}
We will also compare these models to the so-called operational prediction (available from 01/01/09 only, i.e. for the second half of our dataset only) i.e. the final prediction that was actually used by EDF. Note that the operational prediction \(\mathrm{Pred}_{\mathrm{OP}}\) cannot be written as a prediction coming from a statistical model (even though we will sometimes abusively refer to it as the prediction from the operational model) : it combines manual adjustments and statistical models. \(\mathrm{Pred}_{\mathrm{OP}}\) is computed as a 50\%--50\% mixture between the two predictions \(\mathrm{Pred}_{\mathrm{DOAAT}}\) and \(\mathrm{Pred}_{\mathrm{DCo}}\) that we briefly describe below.

The prediction \(\mathrm{Pred}_{\mathrm{DOAAT}}\) is obtained as follows. A model similar to the one described in \cite{Bruhns}, with an ARIMA part, is first used on a real-time estimated signal corresponding to the ''France'' perimeter. An estimated loss is then substracted from it, accounting for the customers within this perimeter that are not affiliated with EDF. A manual adjustment is finally applied in real-time. It is a ``top-down'' prediction in the sense that the ``EDF`` perimeter is approximated as a difference between the ``France`` perimeter and a ''France but not EDF'' perimeter.

The prediction \(\mathrm{Pred}_{\mathrm{DCo}}\) is obtained as follows. Multiple models from \cite{Bruhns} are used upon consolidated signals (not available in real-time, only three weeks later) for sub-perimeters, the reunion of which is the ``EDF`` perimeter. The corresponding predictions are then added together before a manual adjustment is finally applied in real-time. It is a ''bottom-up`` prediction in the sense that the ''EDF`` perimeter is approximated as the sum of all its parts.

There are a number of differences between the dynamic predictions and the operational predictions. First of all, the operational predictions are computed using predicted temperatures (since the sequence of observed temperatures at the time of prediction is clearly not available) whereas the models that we consider (see Section \ref{sec:dynamicmodels}) are based on the realised temperatures. The operational predictions make use of a calendar that includes more daytypes and also benefit from high level expertise through the manual adjustments mentioned. But the biggest difference in nature between these predictions lies somewhere else: the dynamic predictions are made from one day to the next (with no intraday correction whatsoever, since we are basically considering 48 independent models), while the operational predictions are made from one half-hour to the next. Essentially the horizon of prediction for the dynamic models is \(\tau=1\text{ day}=48\text{ half-hours}\) whereas it is much smaller for \(\mathrm{Pred}_{\mathrm{DOAAT}}\), since the new data get incorporated approximately every 8 half-hours (the computation of \(\mathrm{Pred}_{\mathrm{DOAAT}}\) is based upon a real-time signal though, not consolidated data).

\subsection{Results}
\subsubsection{Running the filter}
For the estimation and prediction of the model, we used the Algorithm \ref{algo:filterfinal} with a total number of \(M=10^5\) particles. One time step (filtering and predicting the state with horizon \(\tau = 1\), including 90\% credible intervals) took approximately \(1\) second on a single core Intel(R) Xeon(R) E5410 (2.33GHz) for one of the 48 independent models, which is compatible with the goal of being able to predict the electricity load in an online manner. The execution time grew a bit larger and reached \(3\) seconds per iteration when the predictive horizon was set to \(\tau = 5\). Note that providing credible intervals requires the use of a sorting algorithm, for example quicksort \cite[see][]{Hoare1962} with complexity \(\mathrm{O}(M\log M)\) whereas Algorithm \ref{algo:filterfinal} has only complexity \(\mathrm{O}(M)\). Quicker runtimes are thus obviously achievable if the computation of credible intervals is not needed.

\subsubsection{Degeneracy}
Before looking at the filtered or predicted distributions that we are most interested in, we actually have to assess whether the numerical results obtained are actually usable or not. If the degeneracy problem proved too strong along the estimation process, the estimated values indeed become questionable.

\begin{figure}[htbp]
\begin{center}
\includegraphics[width=1\textwidth]{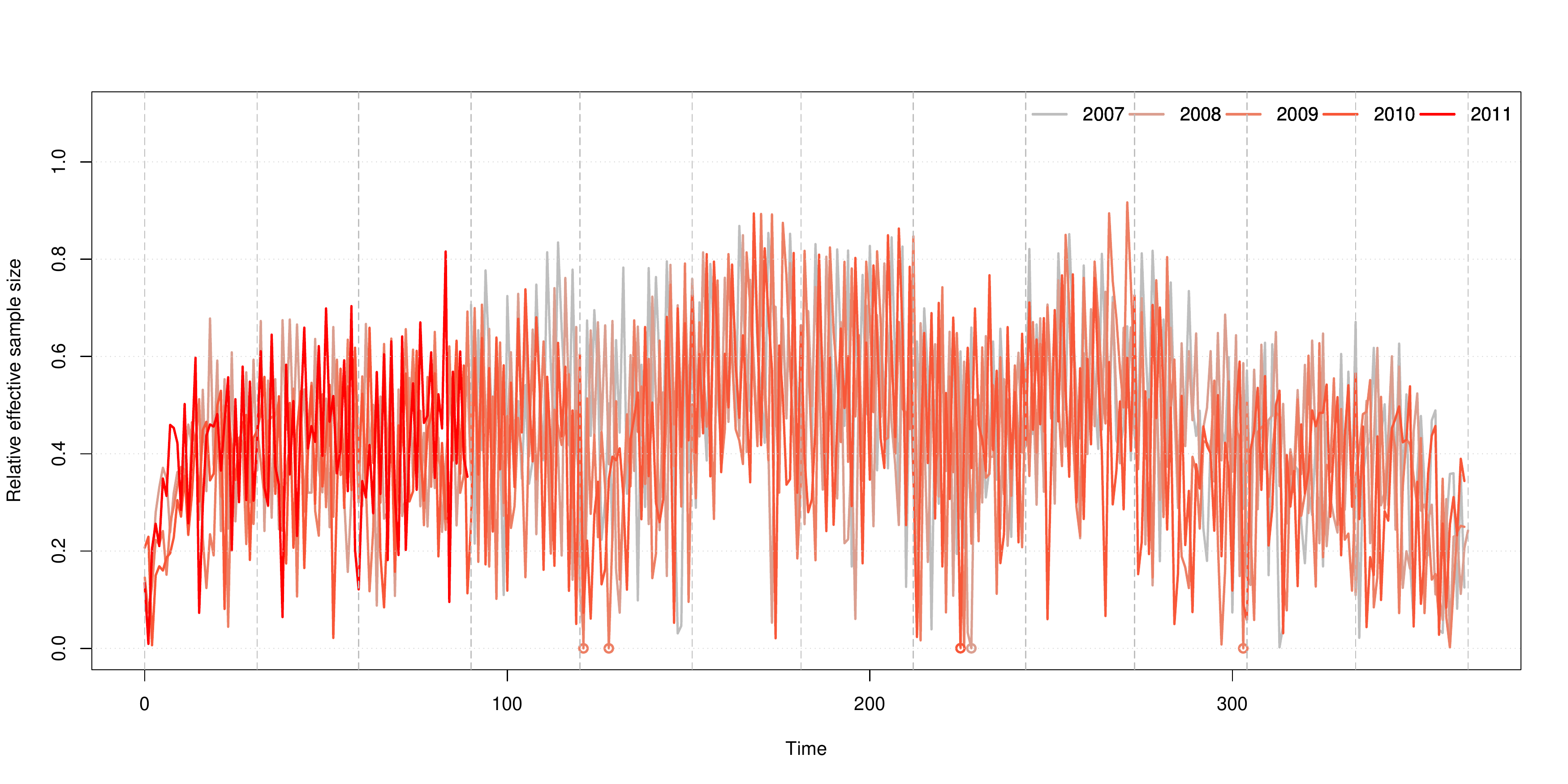}
\end{center}
\caption{Relative effective sample size \(\frac{\mathrm{ESS}(n)}{M}\) for the dynamic model \eqref{eq:maindynamicmodel} at 12:00 as a function of the day in the calendar. The saturation of the colour used increases with each year. Data detected as outliers are marked with a circle.}
\label{fig:essr_24_2layers_horizon5}
\end{figure}

\begin{figure}[htbp]
\begin{center}
\includegraphics[width=1\textwidth]{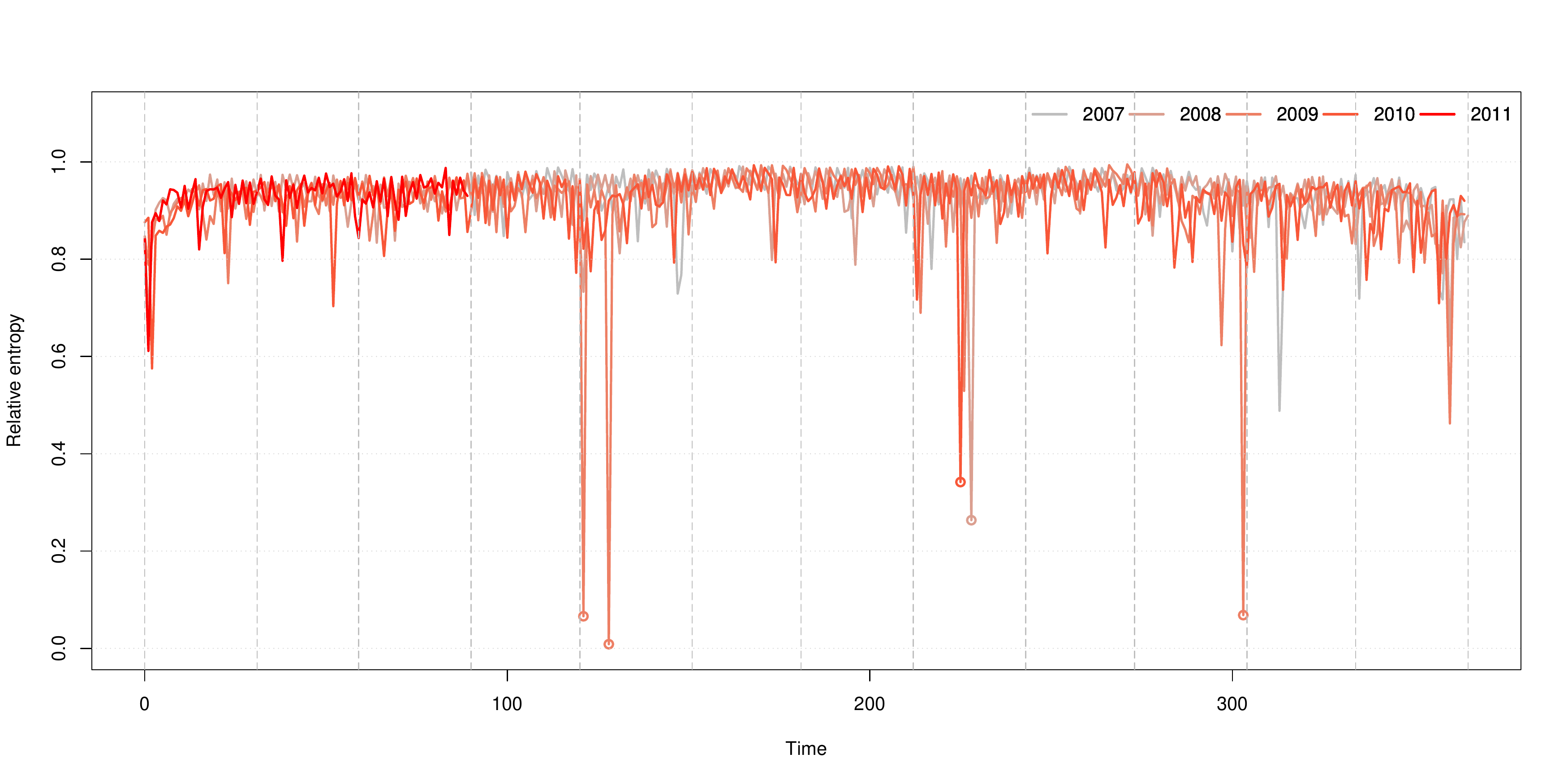}
\end{center}
\caption{Relative entropy \(\frac{\mathcal{E}(n)}{-\log M}\) for the dynamic model \eqref{eq:maindynamicmodel} at 12:00 as a function of the day in the calendar. The saturation of the colour used increases with each year. Data detected as outliers are marked with a circle.}
\label{fig:entropy_24_2layers_horizon5}
\end{figure}

\begin{figure}[htbp]
\begin{center}
\includegraphics[width=1\textwidth]{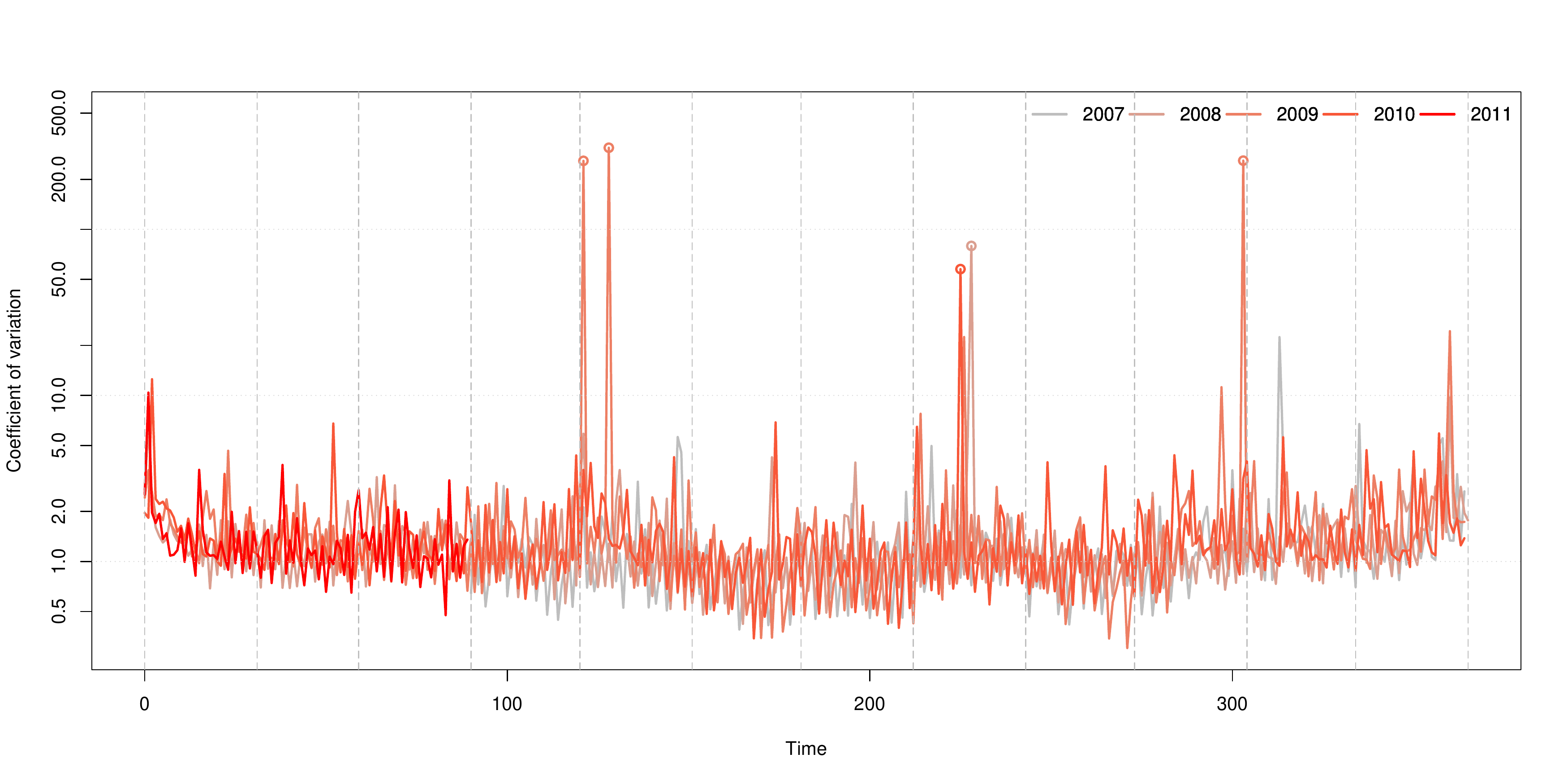}
\end{center}
\caption{Coefficient of variation \(\mathrm{CV}(n)\) for the dynamic model \eqref{eq:maindynamicmodel} at 12:00 as a function of the day in the calendar. The saturation of the colour used increases with each year. Data detected as outliers are marked with a circle. The ordinate axis is in log-scale.}
\label{fig:cv_24_2layers_horizon5}
\end{figure}

Figures \ref{fig:essr_24_2layers_horizon5}, \ref{fig:entropy_24_2layers_horizon5} and \ref{fig:cv_24_2layers_horizon5} show the evolution of the various criteria discussed in Section \ref{sec:monitordegeneracy} throughout time for the model \eqref{eq:maindynamicmodel} at the instant 12:00. These criteria exhibit a seasonal behaviour (with a 1 year period), as the time series itself, showing that the particle filter is subject to a little more degeneracy during winter than during summer (the electricity load is indeed harder to predict, due to the influence of the temperature). Although the coefficient of variation \(\mathrm{CV}(n)\) is only a rescaling of the effective sample size \(\mathrm{ESS}(n)\) (see \eqref{eq:relationESSCV}), the outliers detected by the Algorithm \ref{algo:filterfinal} used are much easier to spot on Figure \ref{fig:cv_24_2layers_horizon5} than on Figure \ref{fig:essr_24_2layers_horizon5}. Also observe that even if the entropy and the coefficient of variation approximate two different divergences \cite[see][for the details]{Cornebise}, the outliers are as easily spotted on Figures \ref{fig:entropy_24_2layers_horizon5} and \ref{fig:cv_24_2layers_horizon5} and the behaviours of the two criteria are very similar : hence, using the entropy instead of the effective sample size (or the coefficient of variation, since they are interchangeable) to detect outliers in Algorithm \ref{algo:filterfinal} could be doable (after having developed a basic intuition of its scaling, in order to decide of a threshold) but would not change the results obtained in any major way.

\subsubsection{Outliers}
\begin{figure}[htbp]
\begin{center}
\includegraphics[width=.49\textwidth]{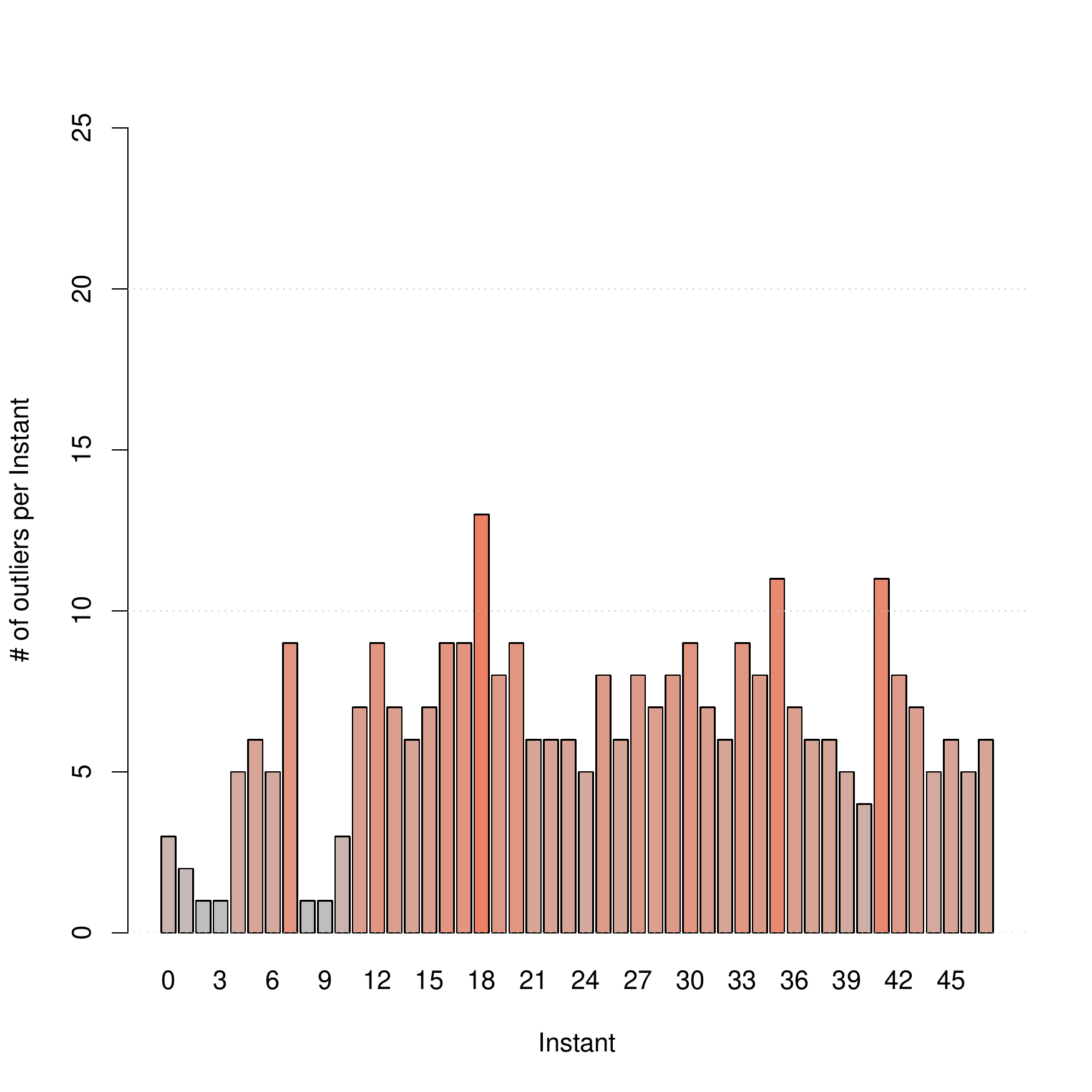}
\end{center}
\caption{Number of outliers detected for each instant of the day by the dynamic model \eqref{eq:maindynamicmodel}.}
\label{fig:skipped_by_instant}
\end{figure}

We show in Figure \ref{fig:skipped_by_instant} the number of data that were automatically detected as outliers  by the model for each instant (half-hour) of the day. Recall that, according to Algorithm \ref{algo:filterfinal}, an outlier is detected whenever the effective sample size would have dropped below 0.1\% of the actual sample size. The amount of outliers varies from one half-hour to the next because an observation flagged as an outlier at a given instant does not necessarily imply that the observation at the next instant will also be flagged. In particular, we observe that more outliers are detected during the day than during the night, which suggests that nighttime is slightly easier to predict than daytime (recall that outliers are essentially data that are badly predicted).

\begin{figure}[htbp]
\begin{center}
\includegraphics[width=1\textwidth]{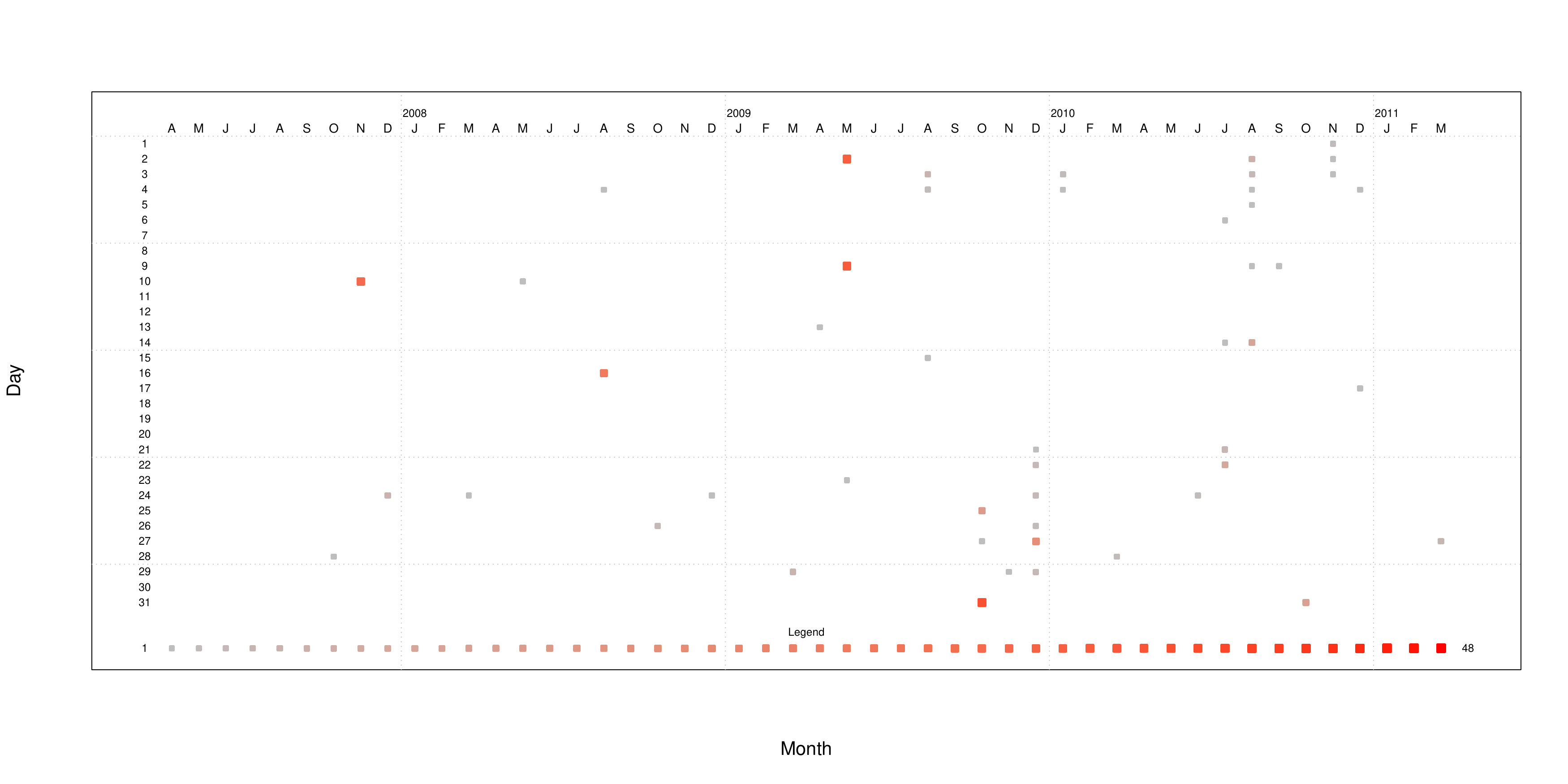}
\end{center}
\caption{Number of outliers detected by the dynamic model \eqref{eq:maindynamicmodel} depending on the calendar from 04/01/2007 to 03/31/2011. Each column represents a month. The size of the point and the saturation of the colour used grow with the number of outliers, as indicated in the legend beneath.}
\label{fig:calendar_skipped}
\end{figure}

Figure \ref{fig:calendar_skipped} shows the number of outliers depending on the calendar. It allows us to pinpoint the times of the year at which these outliers are actually detected. The summer and winter holiday breaks, and the daylight saving time adjustments are easily spotted. Note that for these events, no prior information was available to the dynamic models. Some days before or after bank holidays are also flagged as outliers (05/02, 05/02, 11/10), even though the dynamic model benefits from some calendar information. This should not come as a surprise however: the daytype specification that we chose is rather poor compared to the calendar used for the operational predictions. A more refined calendar, involving specific daytypes, is likely to help turning these few outliers back into regular data, provided the initialisation of the particle filter is correctly done.

Table \ref{tab:cross_kept_dayvalidity} summarises what is already guessable from Figures \ref{fig:calendar_dayvalidity} and \ref{fig:calendar_skipped}, i.e. that most of the (few) instants detected as outliers by the dynamic model are indeed bank-holidays.
\begin{table}[htbp]
\begin{center}
\begin{tabular}{rrr}
  \hline
  instant       & outlier    & not outlier \\
  \hline
  bank-holiday     & 269 [5.60] & 16627 [346.40] \\
  not bank-holiday & 38  [0.79] & 53194 [1108.21] \\
  \hline
\end{tabular}
\end{center}
\caption{Classification of the instants for the dynamic model \eqref{eq:maindynamicmodel}. The number given between square brackets is an equivalent of the number of instants in days (i.e. divided by 48).}
\label{tab:cross_kept_dayvalidity}
\end{table}

\subsubsection{Performance and instants}
We show the overall predictive (horizon \(\tau=1\)) performance of the dynamic model \eqref{eq:maindynamicmodel} against the operational model (OP) in Table \ref{tab:qualite_globale}, depending on whether bank-holidays were included in the calculations or not. The results shown in both cases aggregate the 48 models that were estimated independently from one another. Over the whole period of study, the operational predictions are better than the predictions provided by the dynamic models, but they also do benefit from more specific calendar information being used to compute them. When bank-holidays are removed from the calculations, the overall predictive quality of the dynamic models improves considerably as demonstrated by the results in Table \ref{tab:qualite_globale}.

\begin{table}[htbp]
\begin{center}
\begin{tabular}{rrrr}
  \hline
  & dynamic model
  & operational model \\
  \hline
with bank-holidays    & 1.4342 & 1.2344 \\
without bank-holidays & 1.1712 & 1.2185 \\
  \hline \\
\end{tabular}
\end{center}
\caption{Overall predictive (horizon \(\tau=1\)) and MAPE (in \%) for the dynamic model \eqref{eq:maindynamicmodel} and the operational model. The top row results include bank-holidays in the calculations, while the bottom row results do not.}
\label{tab:qualite_globale}
\end{table}

\begin{figure}[htbp]
\begin{center}
\includegraphics[width=.49\textwidth]{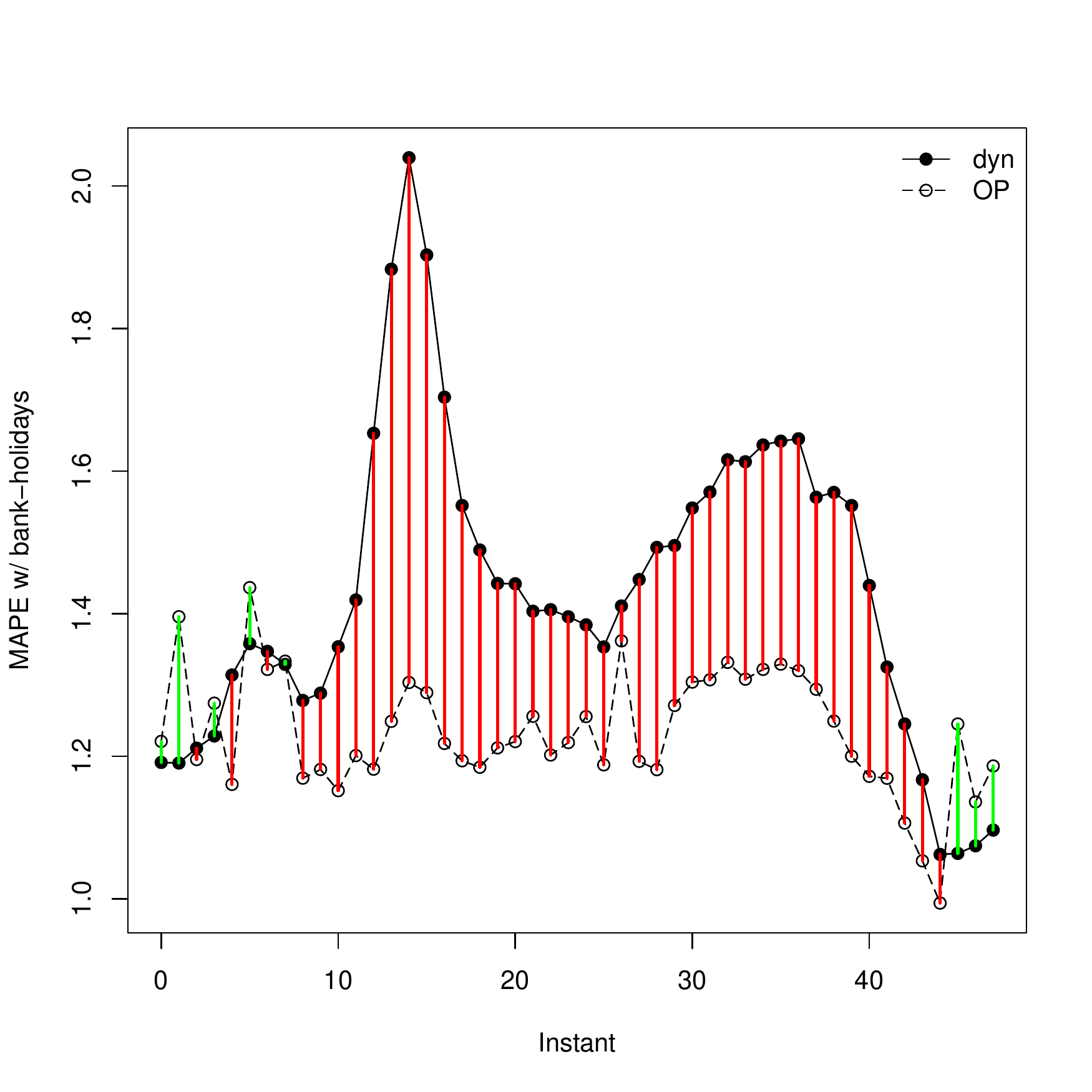}
\includegraphics[width=.49\textwidth]{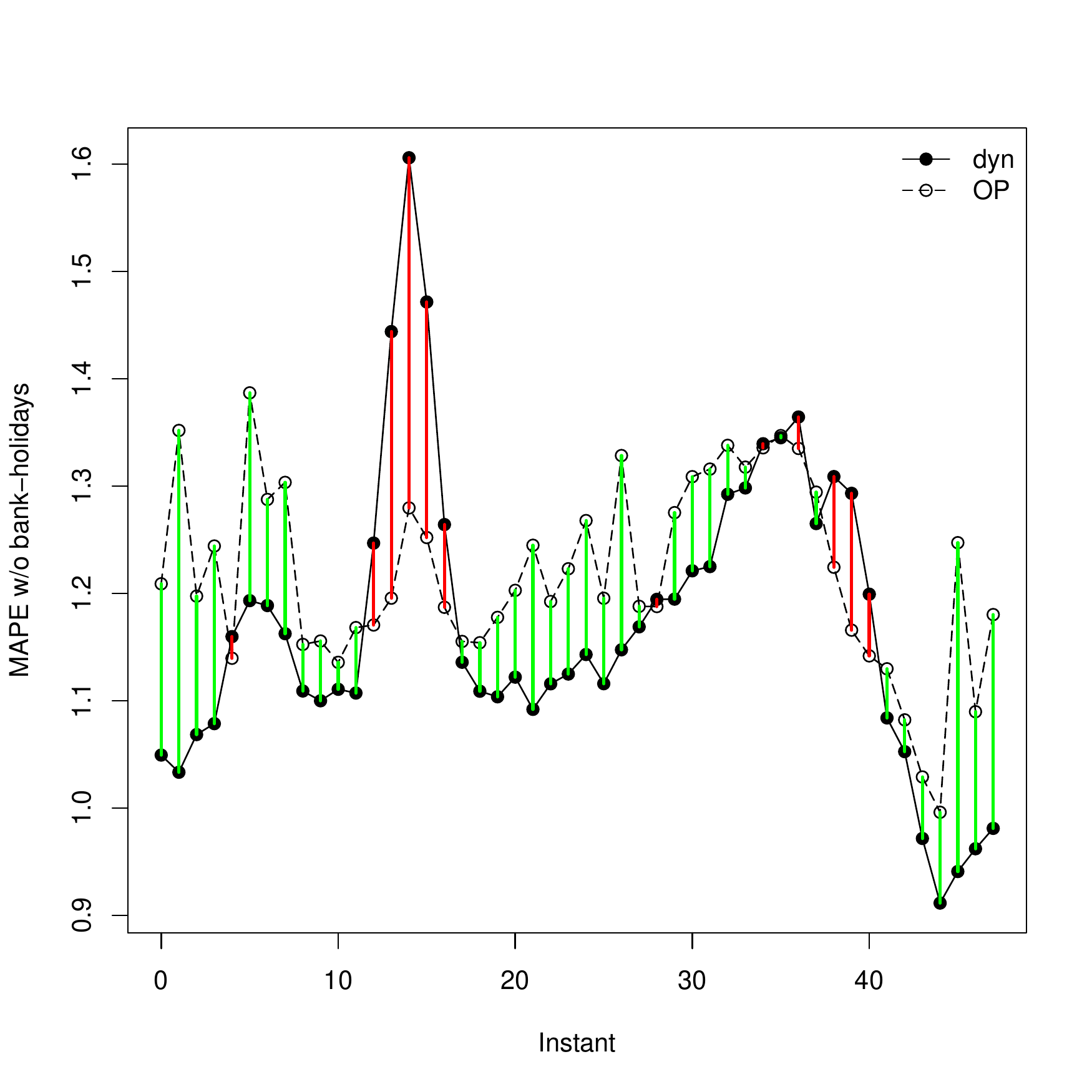}
\end{center}
\caption{Predictive (horizon \(\tau=1\)) and MAPE (in \%) for the dynamic model (dyn) and the operational model (OP) for each of the 48 half-hours, including bank-holidays in the calculations (leftmost figure) and not including bank-holidays in the calculations (rightmost figure). The difference between the two models is coloured depending on its sign: green when the dynamic model is better than the operational model and red when not.}
\label{fig:qualite_instant}
\end{figure}

In fact, looking at Figure \ref{fig:qualite_instant} that represents the predictive MAPE of the dynamic model (dyn) and operational model (OP) averaged by instant, we are able to see that the dynamic model predicts the electricity load quite well when bank-holidays are not considered, challenging the operational model throughout the day, except during the morning ascent. The good predictive performance of the dynamic model on these days is somewhat surprising because the dynamic predictions, coming from 48 independent models, are made from one day to the next whereas the operational predictions include an ARIMA adjustment phase to take advantage of the most recent observations, and also benefit from manual adjustments.

\subsubsection{Performance and horizon}
\begin{figure}[p]
\begin{center}
\includegraphics[width=.49\textwidth]{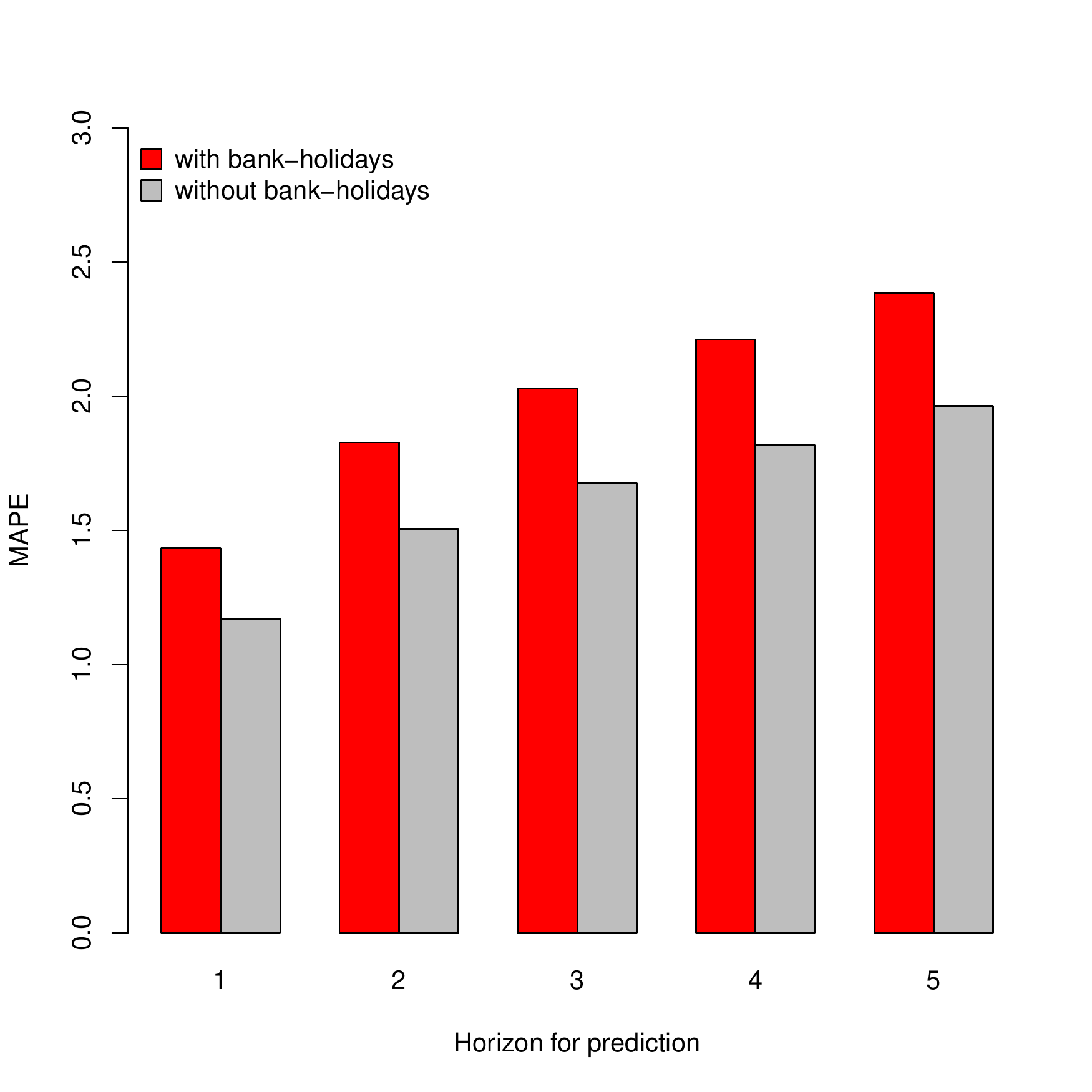}
\end{center}
\caption{Predictive MAPE of the dynamic model \eqref{eq:maindynamicmodel} for \(\tau=1,\ldots,5\), including bank-holidays in the calculations (leftmost bars) and not including bank-holidays in the calculations (rightmost bars).}
\label{fig:qualite_horizon}
\end{figure}

Since the operational predictions are sometimes required up to \(\tau=3\) days, we now investigate the predictive quality of our dynamic model as the horizon for prediction grows larger. Figure \ref{fig:qualite_horizon}, given hereafter, displays the predictive MAPE for horizon \(\tau=1,\ldots,5\), whether including bank-holidays in the calculations or not. It is clear that the predictive errors of the dynamic model increase with the horizon \(\tau\) considered for the prediction, confirming that it is primarily meant for short-term forecasts and not long-term forecasts.

\begin{figure}[p]
\begin{center}
\includegraphics[width=1\textwidth]{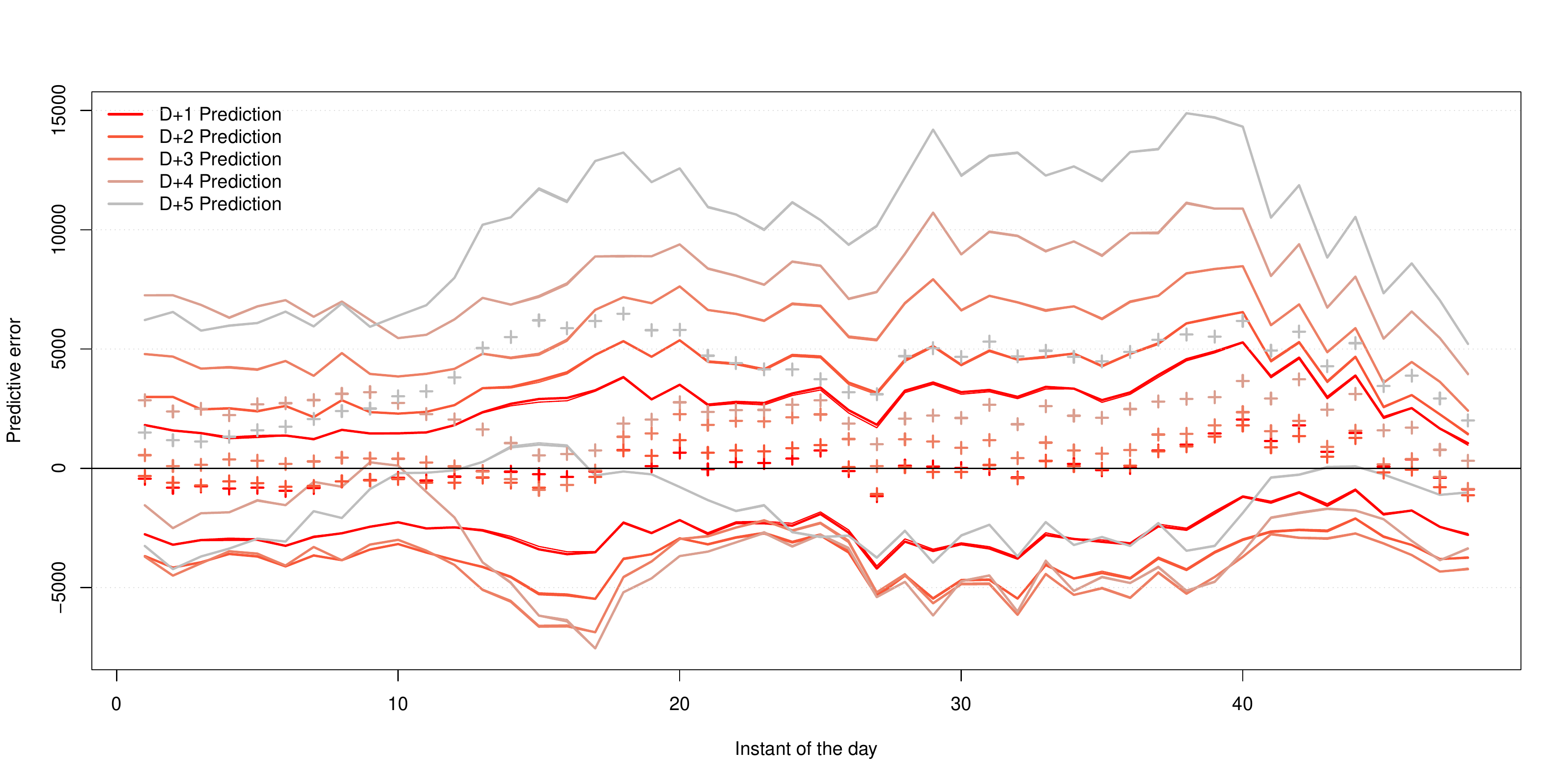}
\end{center}
\caption{Predictive errors (predictive mean minus true value) of the dynamic model \eqref{eq:maindynamicmodel} for the observations on 12/30/2010 (48 half-hours) with \(\tau=1,\ldots,5\). The horizontal black line marks the true load (no predictive error), and crosses mark the various predictive errors with their respective credible intervals in solid lines. The more recent the predictions are (i.e. the smaller \(\tau\) is), the more saturated the colour used is: D+1 Prediction is the most recent prediction (it was made 1 day before) while D+5 is the oldest prediction (it was made 5 days before).}
\label{fig:Prediction_horizon}
\end{figure}

Another consequence of increasing the prediction's horizon is that the credible intervals obtained around the predictions also tend to grow larger on average as can be observed in Table \ref{tab:size_CI_horizon}. An illustration of the credible intervals returned by the dynamic models is given in Figure \ref{fig:Prediction_horizon} where the electricity load is predicted over 48 consecutive instants via the dynamic model \eqref{eq:maindynamicmodel}. The predictions clearly improve over time as the model takes more and more recent information into account: the one-day-ahead predictions about 12/30/2010 provided on 12/29/2010 are much more accurate than the five-days-ahead predictions (of the same day) that were computed on 12/25/2010. Figure \ref{fig:Prediction_horizon} also makes it clear that the credible intervals obtained for a predictive horizon \(\tau=1\) are narrower compared to those obtained for a predictive horizon \(\tau=5\) (but note that their lengths vary over time).

\begin{table}[htbp]
\begin{center}
\begin{tabular}{rrrrrr}
  \hline
  & \(\tau=1\) & \(\tau=2\) & \(\tau=3\) & \(\tau=4\) & \(\tau=5\) \\ \hline
  \(\widehat{\lambda}_{\text{90\%}}({x}_{n+\tau})\) & 2746.3 & 3721.1 & 4505.6 & 5191.6 & 5815.3 \\
  \(\widehat{\lambda}_{\text{90\%}}({y}_{n+\tau})\) & 3036.1 & 3947.7 & 4696.5 & 5358.6 & 5964.9 \\ \hline \\
\end{tabular}
\end{center}
\caption{Mean length (in MW) of the symmetric 90\% credible intervals (CI) around the predicted states \(\widehat{x}_{n+\tau}\) and around the predicted observations \(\widehat{y}_{n+\tau}\) of the dynamic model \eqref{eq:maindynamicmodel}, for \(\tau=1,\ldots,5\).}
\label{tab:size_CI_horizon}
\end{table}

\begin{table}[htbp]
\begin{center}
\begin{tabular}{rrrrrr}
  \hline
  & \(\tau=1\) & \(\tau=2\) & \(\tau=3\) & \(\tau=4\) & \(\tau=5\) \\ \hline
  \(\widehat{\chi}_{\text{90\%}}(\widehat{x}_{n+\tau})\) & 89.569 & 90.442 & 92.385 & 93.479 & 94.168 \\
  \(\widehat{\chi}_{\text{90\%}}(\widehat{y}_{n+\tau})\) & 92.531 & 92.501 & 93.773 & 94.472 & 94.882 \\ \hline \\
\end{tabular}
\end{center}
\caption{Empirical coverage (in \%) of the symmetric 90\% credible intervals (CI) around the predicted states \(\widehat{x}_{n+\tau}\) and around the predicted observations \(\widehat{y}_{n+\tau}\) of the dynamic model \eqref{eq:maindynamicmodel}, for \(\tau=1,\ldots,5\).}
\label{tab:cover_CI_horizon}
\end{table}

The empirical coverages of the symmetric 90\% credible intervals around the predicted states and observations are given in Table \ref{tab:cover_CI_horizon}. These values were computed as the ratio between the number of instants for which the observations fell inside the interval, and the total number of instants. Note that if the observations were mutually independent outcomes of the same random variable (which they are not in our situation because of the exogenous variables temperature and calendar), this ratio would theoretically approximate the true rate of coverage i.e. 90\%. Even so, the empirical coverage computed seems, somewhat reassuringly, to agree with the expected rate.

\subsubsection{Filtered weather parts}
\begin{figure}[p]
\begin{center}
\includegraphics[width=.49\textwidth]{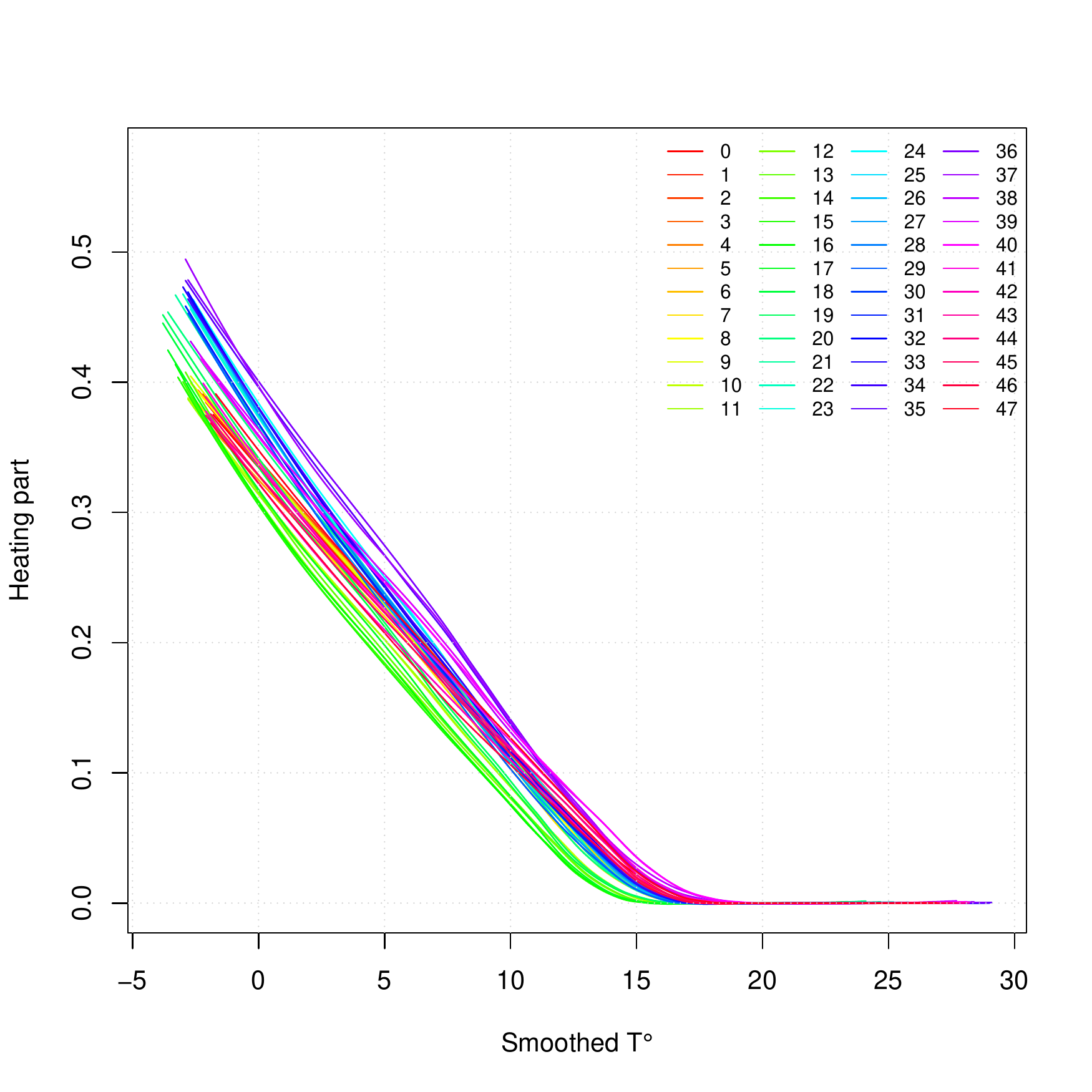}
\end{center}
\caption{Estimated filtered heating part of the dynamic model \eqref{eq:maindynamicmodel} against the smoothed temperature \(T_n^{\text{heat}}\) given to the model. 48 distinct colours are used, one for each of the 48 half-hours. The estimation was done via the loess function in R considering the filtered mean of the heating part against both temperatures. Only the relative heating part is shown here, i.e. the heating part divided by the maximum load observed over the whole period.}
\label{fig:PartChauffagecontreTemperature}
\end{figure}

The Figure \ref{fig:PartChauffagecontreTemperature} shows the filtered heating and cooling parts of the dynamic model \eqref{eq:maindynamicmodel}. It seems to be piecewise linear with regard to the temperature variables upon which it depends, with a threshold that depends on the instant considered. The heating part however is not modelled as such since the heating gradient is chosen non constant in the dynamic model. It is thus a bit of a surprise to find this familiar piecewise linear shape for the heating part, even though it is quite common for non dynamic models \cite[see][for example]{Bruhns}.

As in \cite{Dordonnat}, the heating gradient of the dynamic model appeared to be stronger in winter and slightly weaker over mid-seasons (note that the behaviour of the heating gradient over summer is of little practical importance: while it is true that it cannot be observed accurately at that time, it also has no direct impact on the quality of the model since this is precisely the period over which the heating part of the model vanishes). Note that, on the contrary, the variance of the heating gradient appeared to be larger during summer (with no information available) than during winter.

\subsubsection{Filtered seasonal part}
Even though we do not display it here, let us mention that the filtered seasonal part \(x_n^{\text{season}}\) of the dynamic model exhibits a 1-year period with weekly cycles. Around the main periodic pattern, variations occur : more so over the winter period, for which the seasonal part is obviously not so well defined, than over the summer period. Indeed, during summer the seasonal part is the only active dynamic part of the model, while during winter the heating part also plays an important role : the estimated values of both parts over winter are thus to be interpreted with caution. Still, the filtered seasonal part seems to react correctly to the summer and winter holiday breaks (as we will outline in the next Section), although no particular information was used to flag these time windows for the model.

Because EDF customers now represents a fraction only of the French customers population (instead of the whole), the perimeter of the data varies over time due to customers departures or arrivals (but taking into account that EDF and France perimeters were actually identical until a few years ago, departures are a bit more likely). As a matter of fact, the filtered seasonal part also shows successive yearly drops from 2008 and onwards, which correspond to the financial crisis that arose in late 2008 (and that impacted the French electricity load), or planned customers' departures.

\subsubsection{Summer break}
\begin{figure}[htbp]
\begin{center}
\includegraphics[width=.49\textwidth]{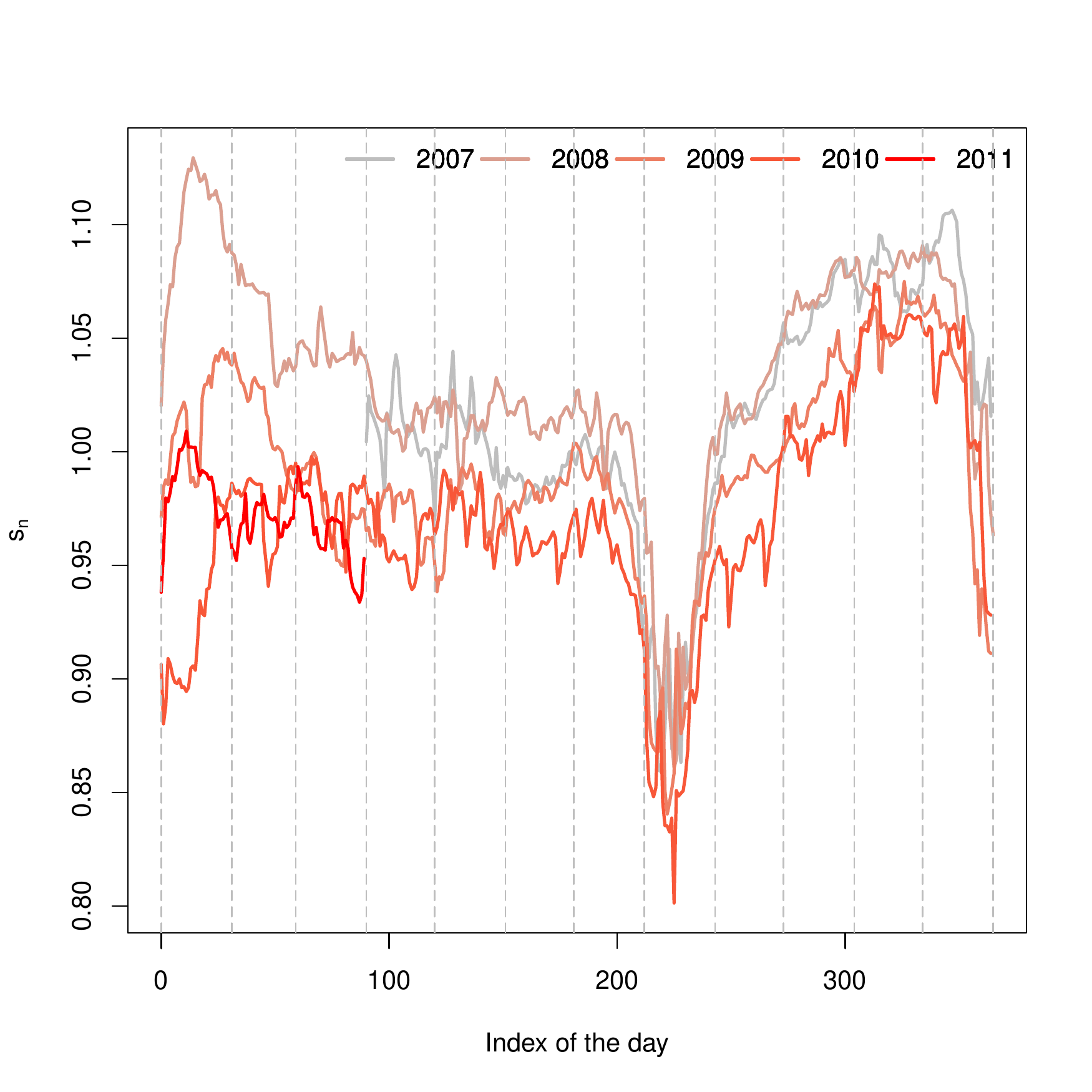}
\includegraphics[width=.49\textwidth]{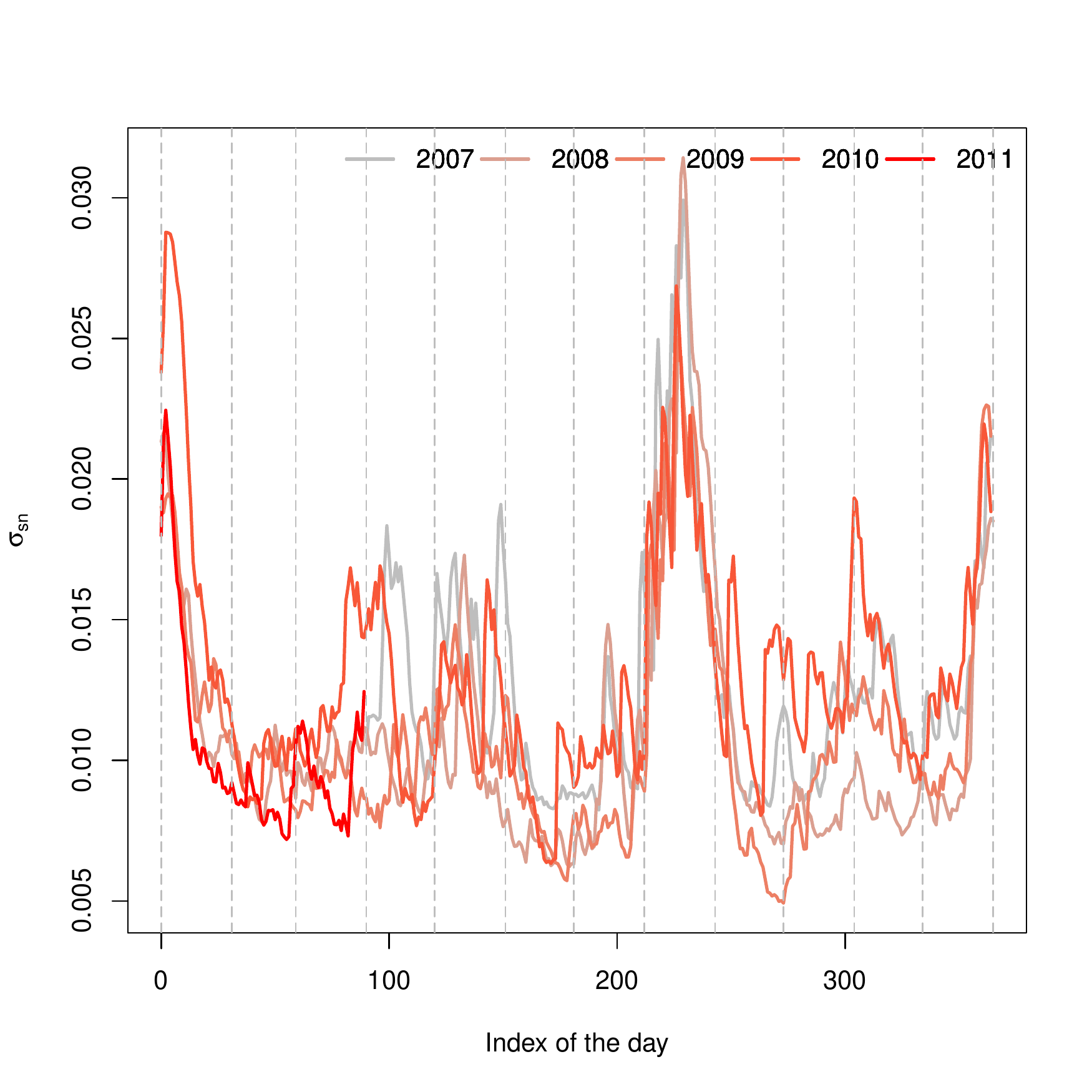}
\end{center}
\caption{Mean of the filtered coefficients \(s_n\) (left) and \(\sigma_{s,n}\) (right) of the dynamic model \eqref{eq:maindynamicmodel} averaged over 48 half-hours, as functions of the day in the calendar. The saturation of the colour used increases with each year. Only the relative filtered means of \(s_n\) and \(\sigma_{s,n}\) are shown here, i.e. the means divided by the mean of \(s_n\) over the whole period.}
\label{fig:summerbreakfiltering}
\end{figure}

Since holiday breaks are among the most toughest times of the year for predictions, we investigate the behaviour of the dynamic models over the summer break to show how the models cope with the difficulty.
\paragraph{Evolution of the dynamics}
The Figure \ref{fig:summerbreakfiltering} shows the filtered mean of both \(s_n\) and \(\sigma_{s,n}\), that rules the dynamic of \(s_n\) within the dynamic model \eqref{eq:maindynamicmodel}. As can be seen on the Figure \ref{fig:summerbreakfiltering}, the model is able to filter out the summer break effectively : to allow for the sharp drop of \(s_n\) during August, the standard deviation of its dynamic \(\sigma_{s,n}\) suddenly grows (becoming twice as large as usual), reflecting the brusque increase of variability of the signal over a short period of time. The model also deals with the winter break in a similar manner.

We have already discussed the behaviour of the heating gradient over summer : during summer the model logically loses track of anything related to the heating part, which leads to artificially increased values of \(\sigma_{g,n}\).

The reasons behind the increased values of \(\sigma_{s,n}\) and \(\sigma_{g,n}\) during the summer break are hence entirely different. Whereas \(\sigma_{s,n}\) grows to allow the model to fit data that do not match the current state, the growth of \(\sigma_{g,n}\) merely reflects the lack of cold temperatures that would help estimate any of the coefficients related to the heating part of the dynamic model.

\paragraph{Predictive errors}
Though no information is provided about the summer break (a succession of breaks mostly occurring on Mondays), we already saw that the dynamic model is able to estimate the electricity load rather correctly given the peculiar circumstances.

A possible way to improve the quality of the forecasts for the days where breaks occur would be to taylor the transition density of one state to the next specifically for them. This requires much expertise in practise because the way the load is affected by the summer break also depends on the calendar configuration: one could for instance introduce adequately modified specifications (interventions) such as
\begin{align*}
s_{n*} = s_{n*-1} - \mu_{n^*} + \epsilon_{n^*}^s
\end{align*}
into the model where \(\mu_{n^*}\in\R_+\) is the drop in load expected to happen at time \(n^*\).

\subsubsection{Comparison with a linear Gaussian state space model}
A dynamic model was proposed and studied by \cite{Dordonnat} to model a similar electricity load series (at the French national perimeter). Their model fit in the multivariate linear Gaussian state space models framework which allowed for the use of Kalman filtering and associated techniques \cite[see][]{Durbin}. It is actually quite a complex and rich model, compared to our own, and includes multiple regressions, some coefficients of which are allowed to vary over time : a truncated Fourier series is used to model the seasonality of the signal as in \cite{Bruhns} in conjunction with a stochastic trend. Local trends are also included to model the holiday breaks, and a calendar with various specific daytypes is used. Heating and cooling parts are defined as well, using fixed threshold values (\(15^\circ\)C and \(18^\circ\)C) as well as fixed smoothing parameters (fixed to \(\vartheta=0.98\)), and are thus very similar to the ones we use, although the heating part relied upon the use of two heating gradients (the first corresponding to the raw temperature, the other to the difference between smoothed and raw temperatures). The model was estimated using national data from 09/01/1995 to 08/30/2004, and its predictive quality was assessed from 09/01/2003 to
08/30/2004 only.

Let us first mention that the performances reported by \cite{Dordonnat} for their model are in accord with ours, with a one-day-ahead predictive MAPE varying around \(1.30\%\) across the 24 hours considered, and larger errors during the weekends or holiday breaks. They also found the quality of the forecasts obtained to be degrading with the predictive horizon, just as we did, and at a similar rate. Finally, the behaviour of the heating gradient that we reported corroborates the behaviour of the heating gradients found in \cite{Dordonnat} (with this difference that they used a smoothing approach for the signal extraction, whereas we used a filtering approach).

Still, the dynamic model \eqref{eq:maindynamicmodel} that we propose is much simpler, most notably where the seasonality part is concerned: our model only includes 9 daytypes and at most 2 temperatures, i.e. 10 random effects whereas the model described in \cite{Dordonnat} made use of more than 30 random effects. Arguably, the estimation time of our model via a particle filter takes more time than running a Kalman filter, but particle filters naturally allows for more flexibility in the definition of the model (including non-linear non-Gaussian model). Most importantly, the Algorithm \ref{algo:filterfinal} that we implemented for the estimation of our models automatically treats bank-holidays instants as missing data when \cite{Dordonnat} explicitly and manually had to declare which data had to be considered as missing data, so as not to throw the model off. Also note that even though the model studied in \cite{Dordonnat} was more complex, the predictive MAPE they obtained for non regular daytypes exceeded \(5\%\) at 09:00AM and 12:00PM the two instants they focused on while the dynamic model \eqref{eq:maindynamicmodel} had an averaged predictive MAPE of \(3.34\%\) for non regular daytypes (but once again keep in mind that the datasets used for their experiments and ours were different which may possibly explain part of the observed difference).


\end{document}